%% file: ECSci.tex
\documentclass[12pt]{article}

\usepackage{scicite}
\usepackage{graphicx}
\usepackage{chapterbib}
\usepackage{bm}%
\usepackage{braket,amsmath}
\usepackage{newfloat}
\DeclareFloatingEnvironment[
    fileext=los,
    name=Figure,
    placement=tbhp,
    within=none,
]{suppfig}

\usepackage{times}
\usepackage[pdftex,bookmarks=true,bookmarksopen,bookmarksnumbered,
                colorlinks,
                linkcolor=blue,
                citecolor=blue]{hyperref}

\newenvironment{sciabstract}{%
\begin{quote} \bf}
{\end{quote}}

\newcounter{lastnote}
\newenvironment{scilastnote}{%
\setcounter{lastnote}{\value{enumiv}}%
\addtocounter{lastnote}{+1}%
\begin{list}%
{\arabic{lastnote}.}
{\setlength{\leftmargin}{.22in}}
{\setlength{\labelsep}{.5em}}}
{\end{list}}

\title{Electrically controlling single spin qubits in a continuous microwave field}

\author
{Arne Laucht,$^{1,\ast}$ Juha T. Muhonen,$^1$ Fahd A. Mohiyaddin,$^1$ \\Rachpon Kalra,$^1$ Juan P. Dehollain,$^1$ Solomon Freer,$^1$ Fay E. Hudson,$^1$ \\Menno Veldhorst,$^1$ Rajib Rahman,$^2$ Gerhard Klimeck,$^2$ \\Kohei M. Itoh,$^3$ David N. Jamieson,$^4$ Jeffrey C. McCallum,$^4$ \\Andrew S. Dzurak,$^1$ and Andrea Morello$^{1,\ast}$\\
\\
\normalsize{$^{1}$Centre for Quantum Computation and Communication Technology, }\\
\normalsize{School of Electrical Engineering and Telecommunications, }\\
\normalsize{UNSW Australia, Sydney, New South Wales 2052, Australia}\\
\normalsize{$^{2}$Network for Computational Nanotechnology, Purdue University, }\\
\normalsize{West Lafayette, Indiana 47907, United States}\\
\normalsize{$^{3}$School of Fundamental Science and Technology, Keio University, }\\
\normalsize{3-14-1 Hiyoshi, 223-8522, Japan}\\
\normalsize{$^{4}$Centre for Quantum Computation and Communication Technology, }\\
\normalsize{School of Physics, University of Melbourne, Melbourne, Victoria 3010, Australia}\\
\\
\normalsize{$^\ast$Corresponding authors. E-mail: a.laucht@unsw.edu.au and a.morello@unsw.edu.au.}
}

\date{}

\begin{document}


\baselineskip24pt


\maketitle


\begin{sciabstract}
Large-scale quantum computers must be built upon quantum bits that are both highly coherent and locally controllable. We demonstrate the quantum control of the electron and the nuclear spin of a single $^{31}$P atom in silicon, using a continuous microwave magnetic field together with nanoscale electrostatic gates. The qubits are tuned into resonance with the microwave field by a local change in electric field, which induces a Stark shift of the qubit energies. This method, known as $A$-gate control, preserves the excellent coherence times and gate fidelities of isolated spins, and can be extended to arbitrarily many qubits without requiring multiple microwave sources.
\end{sciabstract}

The construction of a large-scale quantum computer is among the most exciting scientific challenges of our time. For this purpose, the ability to exploit the advanced fabrication methods developed in semiconductor nanoelectronics \cite{Awschalom2013,Zwanenburg2013} would be highly beneficial. Succeeding at this project will depend upon the ability to create quantum bits (qubits) which are, at the same time, highly coherent and easy to control and couple to each other. However, quantum coherence and ease of control are often antithetic requirements. Several types of semiconductor spin qubits have been designed to be operated using only electric fields \cite{Petta2005,Nowack2007,Medford2013,Kim2014}. This allows control at the nanometer scale via small electrodes, but can lead to unwanted decoherence from charge and gate noise \cite{Dial2013,Laird2013}. Other spin qubits, such as the electrons and nuclei of $^{31}$P donors in silicon \cite{Pla2012,Pla2013} or nitrogen-vacancy (NV) centers in diamond \cite{Bala2008,Maurer2012}, exhibit outstanding coherence times thanks to their lack of sensitivity to electrical noise and the reduced nuclear spin fluctuations in their vicinity. However their coherent operation requires high-frequency oscillating magnetic fields, which are difficult to confine to nanometer scales. A solution to reconcile long coherence with local electrical control was already described in the visionary proposal of Kane \cite{Kane1998} for a donor-based quantum computer in silicon. The resonance frequency of a $^{31}$P donor spin depends on the applied magnetic field $B_0$, as well as the electron-nuclear hyperfine coupling $A$. The latter can be locally tuned with an electrostatic gate, known as the ``$A$-gate'', which deforms the wave function of the donor-bound electron and modifies the probability density at the nucleus~\cite{Rahman2007}. Kane envisaged a multi-qubit quantum computer where a global, always-on microwave magnetic field is by default off-resonance with the qubits. Quantum operations are controlled by locally modifying $A$ in order to bring the desired qubits into resonance with the global microwave field \cite{Wolfowicz2014}. This eminently scalable proposal has been further developed to include quantum error correction \cite{Hollenberg2006}. Here, we present the experimental demonstration of local and coherent electrical control of both the electron and the nuclear spin of a single $^{31}$P donor in silicon, representing the first realization of Kane's $A$-gate on a single qubit.

\begin{figure}[!t]
\begin{center}
\includegraphics[width=0.9\textwidth]{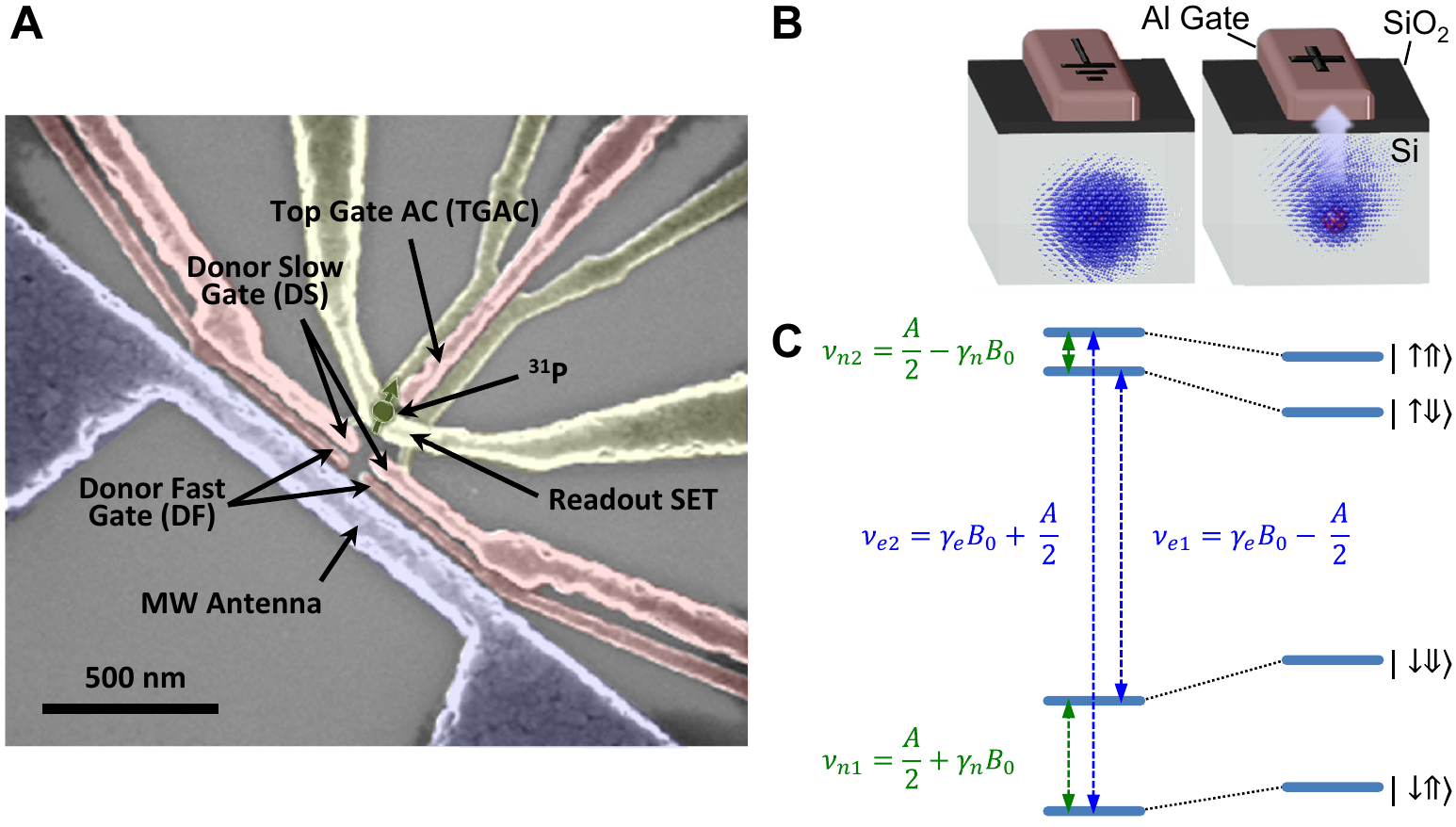}
\caption{\label{figure01} \textbf{Electric field dependence of electron and nuclear energy states.}
\textbf{A}, False-colored scanning electron microscope image of a device similar to the one used in the experiment. Blue: microwave (MW) antenna; Yellow: gates used to induce the single-electron transistor (SET) charge sensor under the SiO$_2$ insulator; Pink: ``A-gates'', comprising gates labeled Donor Fast (DF), Donor Slow (DS) and Top Gate AC (TGAC). These gates are used to tune the potential and electric field at the donor location. \textbf{B}, Electron wavefunction of a donor under an electrostatic gate. A positive voltage applied to the gate attracts the electron towards the Si-SiO$_2$ interface. For illustration purposes, the wavefunction distortion is largely exaggerated as compared to the actual effect taking place in the experiment.
\textbf{C}, Energy level diagram of the neutral e$^{-}$-$^{31}$P system. Gate-controlled distortion of the electron wavefunction modifies $A$ and $\gamma_e$, shifting the electron spin resonance (ESR) $\nu_{e1}$ and $\nu_{e2}$, and the nuclear magnetic resonance (NMR) $\nu_{n1}$ and $\nu_{n2}$ transition frequencies.}
\end{center}
\end{figure}

\begin{figure}[!t]
\begin{center}
\includegraphics[width=0.5\textwidth]{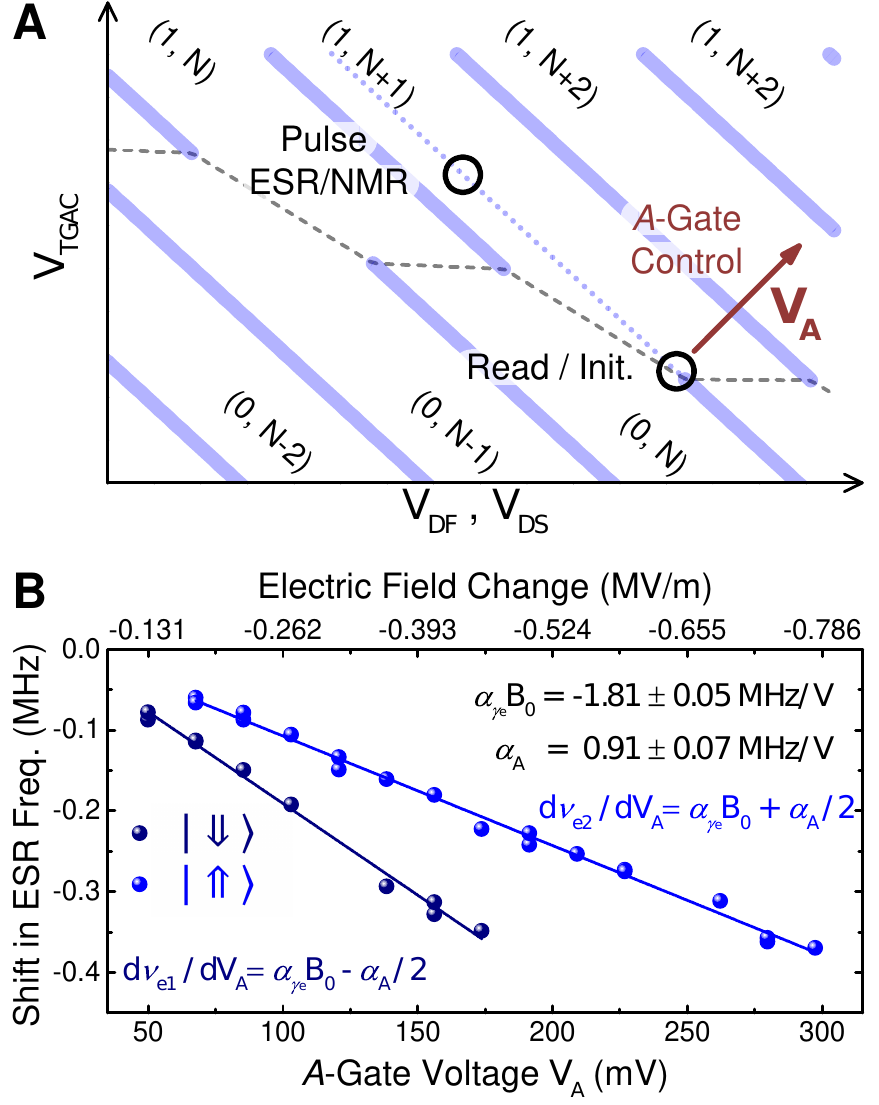}
\caption{\label{figure02} \textbf{Local electrical control of the ESR transition frequencies.}
\textbf{A}, Schematic of the charge stability diagram for this device. The thick solid lines represent the Coulomb peaks of the SET, while the dashed line indicates the ionization/neutralization of the donor.
\textbf{B}, Measured shift in ESR frequencies $\nu_{e1,2}(V_{A})$ as a function of the $A$-gate voltage $V_{A}$. Accurate values of $\nu_{e1,2}(V_{A})$ are obtained by coherent Ramsey experiments (see text). The change in local electric field is obtained from finite-element electrostatic modeling for the specific device geometry and donor location (see supplementary section \ref{TCAD} for details).}
\end{center}
\end{figure}

The $^{31}$P donor in silicon constitutes a two-qubit system, where both the electron (indicated with $\ket{\downarrow}$ or $\ket{\uparrow}$) and the nuclear ($\ket{\Downarrow}$ or $\ket{\Uparrow}$) spin states can be coherently controlled by a magnetic field $B_1$ oscillating at specific electron spin resonance (ESR) and nuclear magnetic resonance (NMR) frequencies. We fabricated a device that comprises a single $^{31}$P donor in an isotopically purified $^{28}$Si epilayer~\cite{Itoh2014}, implanted \cite{Jamieson2005} next to the island of a single-electron-transistor (SET). The SET is formed under an 8~nm thick SiO$_2$ layer by biasing a set of electrostatic gates (yellow in Fig.~\ref{figure01}A). The distance between donor and SET island is $\sim 20(5)$~nm, resulting in a tunnel coupling of order 10~kHz. The device is cooled by a dilution refrigerator (electron temperature $T_{\rm el} \approx 100$~mK), and subject to a static magnetic field $B_0 = 1.55$~T applied along the [110] Si crystal axis. Due to the Zeeman effect the electrochemical potential $\mu$ of the donor electron depends on its spin state, with $\mu_{\uparrow} > \mu_{\downarrow}$. Another set of gates (pink in Fig.~\ref{figure01}A) is used to tune the electrochemical potentials of donor and SET island ($\mu_{\rm SET}$) to the readout position (``Read/Init.'' in Fig.~\ref{figure02}A), where $\mu_{\uparrow} > \mu_{\rm SET} > \mu_{\downarrow}$ and only the $\ket{\uparrow}$ state can tunnel out of the donor. The positive donor charge left behind shifts the SET bias point and causes a current to flow, until a $\ket{\downarrow}$ electron tunnels back onto the donor. This spin-dependent tunneling mechanism is, therefore, used to achieve single-shot electron spin readout \cite{Morello2009,Morello2010}, as well as $\ket{\downarrow}$ initialization. For coherent spin control, an oscillating magnetic field $B_1$ is delivered to the donor by an on-chip, broadband transmission line terminating in a short-circuited nanoscale antenna~\cite{Dehollain2013} (blue in Fig.~\ref{figure01}A). While manipulating the electron spin state, the gates are tuned such that $\mu_{\uparrow,\downarrow} < \mu_{\rm SET}$, to ensure that the electron cannot escape the donor (``Pulse ESR/NMR'' and ``$A$-Gate Control'' positions in Fig.~\ref{figure02}A).

In the present experiment, the same gates used to tune the donor to the readout position are also used to apply an electric field shift, which in turn modifies the spin transition frequencies via the Stark effect. Therefore, we label them ``$A$-gates'' henceforth. In Fig.~\ref{figure01}B we illustrate the effect of the electric field on the donor electron. A positive bias on a gate located above the donor pulls the electron wave function towards the Si/SiO$_2$ interface and away from the nucleus \cite{Rahman2007}. This modifies both the hyperfine coupling $A$ ($\approx 117.53$~MHz in bulk, and 96.9 MHz for this device) \cite{Rahman2007,Muhonen2014} and the electron gyromagnetic ratio $\gamma_e$ ($\approx 27.97$~GHz/T in bulk) \cite{Rahman2009}. The ESR frequencies $\nu_{e1,2}$ depend on both these parameters, and on the state of the $^{31}$P nuclear spin. In the limit $\gamma_e B_0 \gg A$, we find $\nu_{e1} = \gamma_e B_0 - A/2$ (for nuclear spin $|{\Downarrow\rangle}$) and $\nu_{e2} = \gamma_e B_0 + A/2$ (for $|{\Uparrow\rangle}$), as shown in Fig.~\ref{figure01}C. Identifying whether the instantaneous ESR frequency is $\nu_{e1}$ or $\nu_{e2}$ constitutes a single-shot, quantum nondemolition nuclear spin readout \cite{Pla2013}. The NMR transitions can be gate-tuned as well, albeit only via the Stark-shift of $A$, since $\nu_{n1} = A/2 + \gamma_n B_0$ (for electron spin $|{\downarrow\rangle}$) and $\nu_{n2} = A/2 - \gamma_n B_0$ (for $|{\uparrow\rangle}$), where $\gamma_n = 17.23$~MHz/T is the $^{31}$P nuclear gyromagnetic ratio.

During pulse-ESR/NMR experiments we normally operate the electrostatic gates in a compensated manner, to keep $\mu_{\rm SET}$ constant while shifting $\mu_{\uparrow,\downarrow}$ with respect to it \cite{Morello2009} (``Pulse ESR/NMR'' position in Fig.~\ref{figure02}A). This, however, results in a limited variation of the electric field at the donor site. To induce a significant Stark shift we adopted an uncompensated gating scheme, where both $\mu_{\uparrow,\downarrow}$ and $\mu_{\rm SET}$ are drastically lowered (``$A$-Gate Control'' position in Fig.~\ref{figure02}A, see supplementary section \ref{FreqResp} for details on the exact gate configuration). The resulting change in electric field can be calculated with a finite-element Poisson equation solver (TCAD) for the specific device geometry and the triangulated donor location (see supplementary sections \ref{Triag} and \ref{TCAD} for details)~\cite{Mohiyaddin2013}.

We demonstrate the gate-induced shift of the ESR frequencies by performing, with conventional pulse-ESR, a series of Ramsey experiments on the electron spin at different values of $V_{A}$ (Fig.~\ref{figure02}B), where we define $V_A = 0$ as the readout position. The spin is rotated from $\ket{\downarrow}$ to the $xy$-plane by a $\pi/2$ pulse at frequency $\nu_{\rm MW}$, then left to freely precess for a time $\tau$, then rotated by $\pi/2$ again. The accumulation of a phase shift between the spin precession at $\nu_e(V_A)$ and the MW reference clock at $\nu_{\rm MW}$ gives rise to oscillations in the probability of finding the electron $\ket{\uparrow}$ at the end of the sequence. The frequency of the Ramsey fringes gives a very accurate value for $\nu_e(V_A) - \nu_{\rm MW}$.

Both $\nu_{e1}$ and $\nu_{e2}$ shift to lower frequencies upon increasing $V_A$ (Fig.~\ref{figure02}B). This indicates that a significant Stark shift of $\gamma_e$ (i.e. the electron $g$-factor) is taking place, in addition to the $A$-shift. Linear fits to $\nu_{e1,2}(V_A)$ yield slopes
$d\nu_{e1}/dV_A = -2.27(6)$~MHz/V and
$d\nu_{e2}/dV_A = -1.36(3)$ MHz/V.
Using the expressions given in Fig.~\ref{figure02}B we extract the tuning parameters
$\alpha_{A} = dA/dV_A = d\nu_{e2}/dV_A - d\nu_{e1}/dV_A = 0.91(7)$~MHz/V and
$\alpha_{\gamma_e} B_0 = d\gamma_e B_0/dV_A= (d\nu_{e2}/dV_A + d\nu_{e1}/dV_A)/2 = -1.81(5)$~MHz/V, with
$\alpha_{\gamma_e} = d\gamma_e/dV_A = -1.17(3)$~MHz/V/T at $B_0 = 1.55$~T. The positive value of $\alpha_{A}$ indicates that increasing $V_A$ leads to an increase in the electron probability density at the nucleus. This is because a strong electric field is already present for the purpose of forming the SET, whereas increasing $V_A$ causes an additional electric field in the opposite direction, thus an overall reduction of the hyperfine Stark shift (see finite element simulations in supplementary section \ref{TCAD}). The absolute value of the hyperfine coupling and its tunability is in agreement with atomistic tight-binding simulations with NEMO-3D~\cite{Klimeck2007,Klimeck2007b} for the range of donor positions, electric fields, and strain expected for this device (see supplementary section \ref{HFsim}).

An $A$-gate voltage $V_A \sim 300$~meV results in a frequency tuning range $\Delta\nu_{e}^{\rm max} = 400-700$~kHz. Thanks to the use of an isotopically enriched $^{28}$Si epilayer \cite{Itoh2014}, this is a factor $>200$ larger than the intrinsic linewidth $\delta\nu_{e}^{\rm FWHM} = 1.8$~kHz of the ESR transitions~\cite{Muhonen2014}. For the same $V_A$, the NMR frequencies can be shifted by $\Delta\nu_{n}^{\rm max} = 125-150$~kHz by the Stark shift of $A$ alone, which is a factor $\sim250$ higher than the intrinsic linewidth $\delta\nu_{n}^{\rm FWHM} = 0.5$~kHz~\cite{Muhonen2014}.


\begin{figure*}[!t]
\begin{center}
\includegraphics[width=1\textwidth]{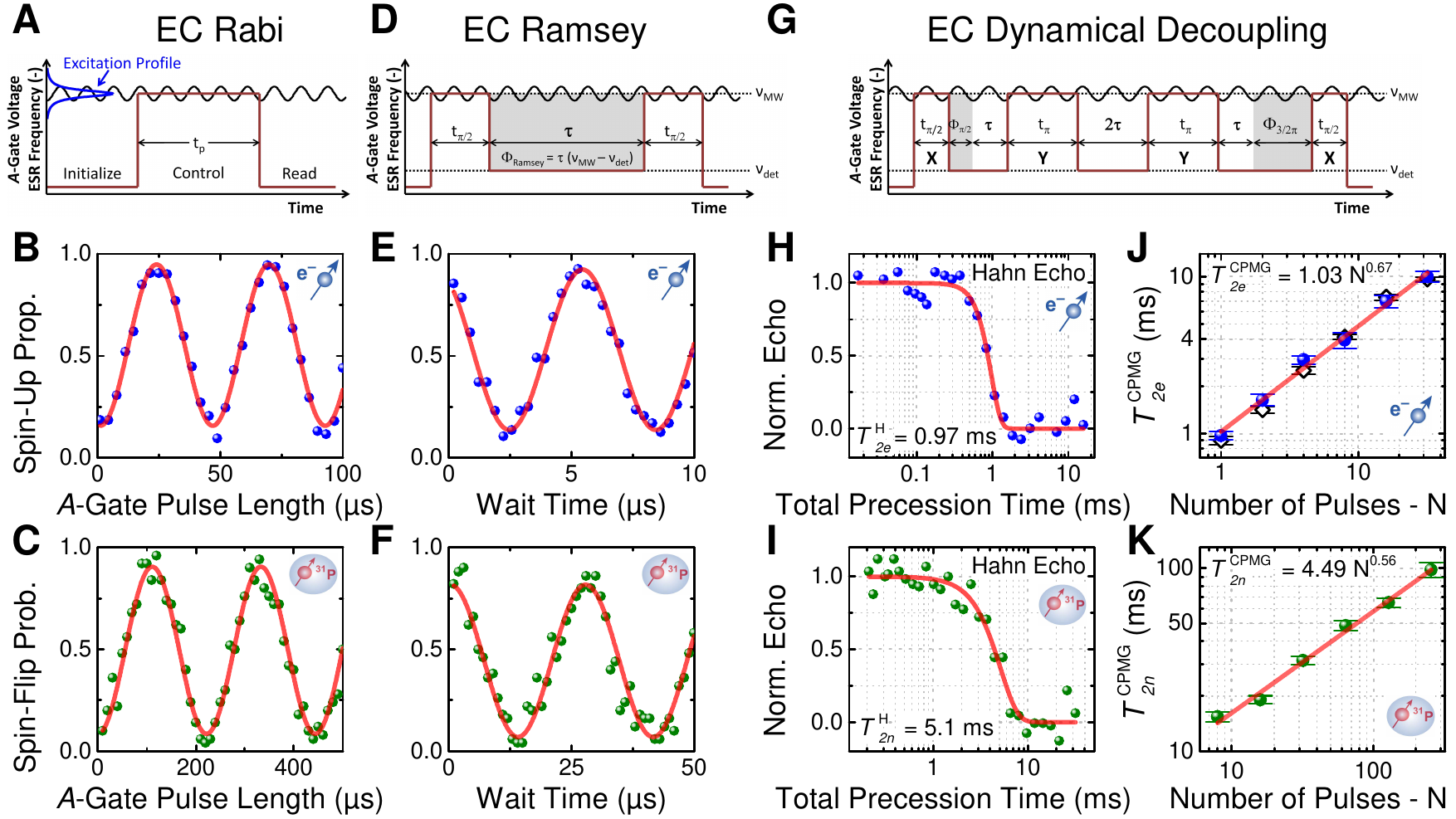}
\caption{\label{figure03} \textbf{Electrically-controlled qubit control and coherence measurements.}
\textbf{A}, Schematic of the sequence used to measure EC Rabi oscillations.
\textbf{B, C}, EC Rabi oscillations measured on the $^{31}$P electron and nucleus, respectively.
\textbf{D}, Schematic of the sequence used to measure EC Ramsey oscillations.
\textbf{E, F}, EC Ramsey oscillations measured on the $^{31}$P electron and nucleus.
\textbf{G}, Schematic of the sequence used to measure EC coherence times.
\textbf{H, I}, EC Hahn echo decay for the $^{31}$P electron and nucleus.
\textbf{J, K}, Extended spin coherence times $T_2$ for CPMG dynamical decoupling sequences on the $^{31}$P electron and nucleus.}
\end{center}
\end{figure*}

Having calibrated the voltage-controlled qubit frequency shifts, we demonstrate how to use $A$-gate pulses to perform coherent control of the qubit states around the Bloch sphere in the presence of a continuous-wave (CW) oscillating magnetic field $B_1$. We demonstrate this on both the electron and the nuclear spin qubits. Fig.~\ref{figure03}A is a schematic of the electrically controlled Rabi sequence. We pulse $V_{A}$ to $V_r$, the voltage needed to tune the spin transition in resonance with the MW or RF source, for a time $t_{p}$, to coherently drive the spin around the $X$-axis of the Bloch sphere, defined in the reference frame rotating at $\nu_{\rm MW}$ (for ESR) or $\nu_{\rm RF}$ (for NMR). The coherent Rabi oscillations (Fig.~\ref{figure03}B,C) are used to calibrate the duration of the control pulses for any desired rotation angle.

An electrically-controlled Ramsey experiment (Fig.~\ref{figure03}D) is obtained by tuning the spin transition frequencies into resonance for the duration of a $\pi/2$-rotation, then moving them to a detuned value $\nu_{\rm det}$ for a wait time $\tau$ to accumulate a phase shift $\Phi_{R}=\tau (\nu(V_r) - \nu_{\rm det})$ with respect to the $\nu(V_r)$ reference frame, and finally tuning them back into resonance for the second $\pi/2$-rotation (Fig.~\ref{figure03}E,F). Qubit rotations around the $Y$-axis can be achieved by accumulating an additional $\Phi_{R} = \pi/2$ before bringing the spin into resonance. This allows for full two-axis ($X$ and $Y$) control of the qubits' states on the Bloch sphere. Alternatively, the phase $\Phi_{R}$ accumulated while off-resonance can be used to produce a controllable $Z$-gate.

We measure the qubits' coherence times $T_2$ by performing Hahn echo and Carr-Purcell-Meiboom-Gill (CPMG) dynamical decoupling sequences (Fig.~\ref{figure03}G-K). For these measurements the wait times $\tau$ are chosen such that they always result in a multiple of $2\pi$ phase shift. The Hahn echo yields $T^{\rm H}_{2e}=0.97(6)$~ms and $T^{\rm H}_{2n}=5.1(3)$~ms for the donor electron and nuclear spins, respectively (Fig.~\ref{figure03}H,I). A CPMG sequence (see schematic in Fig.~\ref{figure03}G for details) further decouples the qubits from low-frequency noise, and extends the coherence times to up to $T^{\rm CPMG}_{2e}=10.0(8)$~ms and $T^{\rm CPMG}_{2n}=98(9)$~ms by applying 32 and 256 $\pi$-pulses, respectively (Fig.~\ref{figure03}J,K). The coherence times of the electrically-controlled electron qubit, measured in milliseconds, can be fitted by $T^{\rm CPMG}_{2e}(N) = 1.03 N^{0.67}$. Conventional pulse-ESR operation (black diamonds in Fig.~\ref{figure03}J) gives nearly identical values, $T^{\rm pulse CPMG}_{2e} = 0.93 N^{0.70}$ (see also supplementary section \ref{DD} for more data). The electrical control method preserves the excellent spin coherence of the qubits, because no additional coupling between the spin states and the environment has been introduced.

\begin{figure}[!t]
\begin{center}
\includegraphics[width=0.5\textwidth]{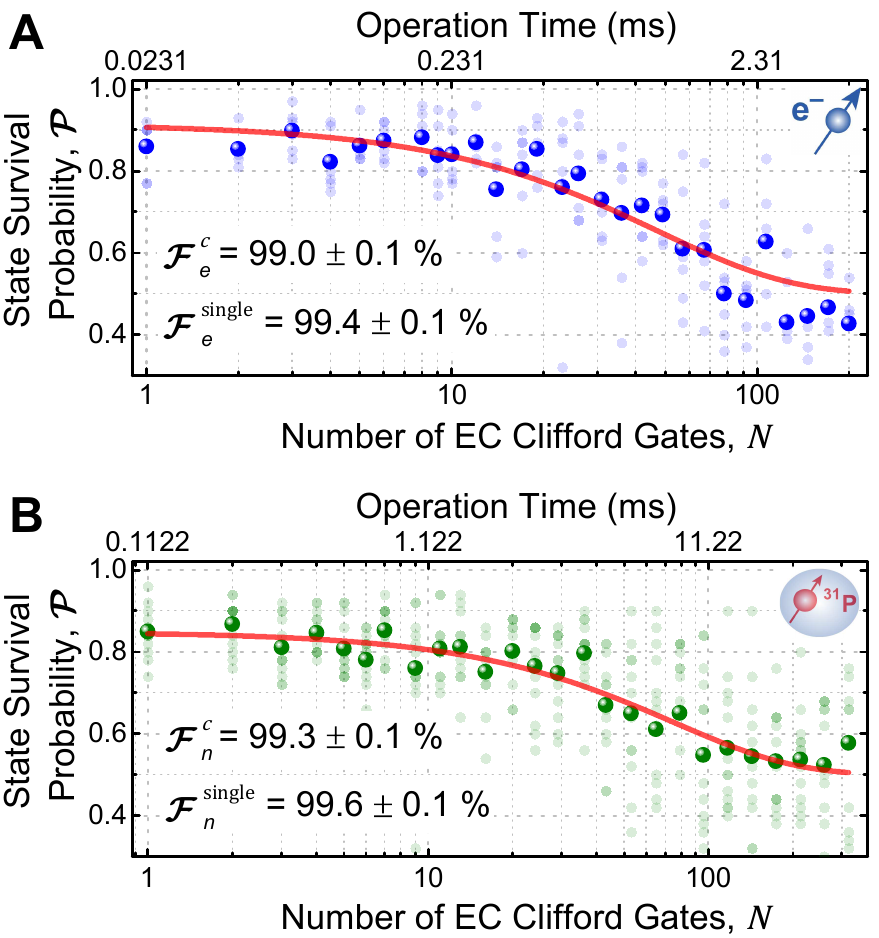}
\caption{\label{figure04} \textbf{Electrically controlled gate fidelities.}
\textbf{A, B} EC randomized benchmarking performed on the $^{31}$P electron and nucleus, respectively. Shaded circles are the results of individual measurements (i.e. individual random sequences of gate operations), while the solid circles show the average state survival probability of all random sequences with the same number of gate operations.}
\end{center}
\end{figure}

We apply the two-axis control to construct the complete set of 1-qubit Clifford gates, and conduct randomized benchmarking~\cite{Knill2008,Muhonen2015} experiments (Fig.~\ref{figure04}A,B) to quantify the average gate fidelities using electrically-controlled qubit manipulation (Fig.~\ref{figure04}). The method relies upon the measurement of the probability $\mathcal{P}(N)$ of arriving at the correct final qubit state after performing a long sequence of randomly-chosen quantum gates. In the presence of gate errors, the probability decays according to the number of operations $N$. We then extract the average fidelity for a single Clifford gate $\mathcal{F}^c$ by fitting the equation
\begin{equation}
\mathcal{P}(N) =M (2\mathcal{F}^c-1)^N + 0.5
\end{equation}
to the data~\cite{Muhonen2015}. $M$ is a free parameter, and depends on the initialization and readout fidelity. We obtain $\mathcal{F}^{c}_{e} = 99.0(1)$~\% and $\mathcal{F}^{c}_{n} = 99.3(1)$~\% for the electron and the nuclear spin, respectively. Since each Clifford gate is composed on average of 1.875 individual gate operations \cite{Epstein2013,Muhonen2015}, we can also quote average single-gate fidelities of $\mathcal{F}^{\rm single}_{e} = 99.4(1)$~\% for the electron and $\mathcal{F}^{\rm single}_{n} = 99.6(1)$~\% for the nucleus. These gate fidelities are comparable to those obtained by pulse-ESR/NMR randomized benchmarking for similar microwave powers~\cite{Muhonen2015}, and are mostly limited by the ratio of gate time to coherence time.

The speed of an electrically-controlled gate operation is inherently limited by the voltage tunability of the resonance frequencies. The excitation profile of the CW field (Fig.~\ref{figure03}A) must be narrow enough to leave the qubits unperturbed while off-resonance, imposing the condition $\gamma B_1 \ll \Delta \nu$. For linearly oscillating $B_1$, $(\gamma B_1)^{-1}$ is the duration of a $\pi$-rotation. Therefore, $\Delta \nu < 1$~MHz in this $^{31}$P device requires gate times $> 10$~$\mu$s.


The qubit control method demonstrated here is applicable to any resonantly-driven qubit where the resonance frequency $\nu_r$ can be quickly and locally controlled by an electric field, and shifted by much more than the resonance linewidth. A wide variety of qubits can potentially fulfil this condition. For spins in diamond \cite{Dolde2011} and silicon carbide \cite{Falk2014} $\nu_r$ can be tuned by modifying the crystal field parameters in the spin Hamiltonian. Magnetic molecules can have tunable $\nu_r$ through a hyperfine Stark effect \cite{Thiele2014} similar to the one shown here. Several types of semiconductor quantum dot qubits exhibit tunable electron spin $g$-factor \cite{Schroer2011,Veldhorst2014} through spin-orbit coupling effects, or tunable splitting through the interplay of valley-orbit and tunnel couplings \cite{Kim2014}.

Due to the high cost of vector microwave signal generators, it seems implausible that future multi-qubit experiments will resort to a dedicated source for each qubit. Time- and frequency-multiplexing qubit control is possible in proof-of-principle experiments, but is impractical in large fault-tolerant quantum processors. In most error correction schemes, fault tolerance is only guaranteed if all qubits can be operated simultaneously at any time. The method of qubit control demonstrated here fulfills all the practical requirements for a large-scale quantum computer, since control gates can be applied simultaneously to arbitrarily many qubits, while requiring only one CW microwave source together with inexpensive multi-channel baseband pulse generators.

\bibliography{Papers}

\begin{scilastnote}
\item[Acknowledgements:] We thank S. Simmons, G. Tosi and C. D. Hill for fruitful discussions and comments. This research was funded by the Australian Research Council Centre of Excellence for Quantum Computation and Communication Technology (project number CE110001027) and the US Army Research Office (W911NF-13-1-0024). We acknowledge support from the Australian National Fabrication Facility, and from the laboratory of Prof. Robert Elliman at the Australian National University for the ion implantation facilities. The work at Keio has been supported in part by FIRST, the Core-to-Core Program by JSPS, and the Grant-in-Aid for Scientific Research and Project for Developing Innovation Systems by MEXT. NCN/nanohub.org computational resources funded by the National Science Foundation under contract number EEC-1227110 were used in this work.
 
\item[Supplementary Information]accompanies the paper.

\end{scilastnote}

\include{suppinf}

\end{document}

%% file: suppinf.tex
\newpage
\bibliographystyle{Science}

\section*{Supplementary Information}

\renewcommand\thesubsection{S\arabic{subsection}}
\setcounter{page}{1}
\renewcommand*{\thepage}{S\arabic{page}}
\renewcommand\thesuppfig{S\arabic{suppfig}}

\subsection{Device Fabrication}
The device was fabricated on a $0.9$~$\mu$m thick epilayer of isotopically purified $^{28}$Si, grown on top of a $500$~$\mu$m thick
$^{\rm nat}$Si wafer. The $^{29}$Si has been depleted to $800$~ppm in the enriched $^{28}$Si epilayer. Single-atom qubits were selected out of a small group of donors implanted in a region adjacent to the Single-Electron-Transistor (SET). In this device, P$^+_2$ molecular ions were implanted at $20$~keV energy in a $100\times100$~nm$^2$ window.
All other nanofabrication processes were identical to those described in detail in Ref.~\cite{Pla2012}, except for a slight modification in the gate layout to bring the qubits closer to the microwave antenna and provide an expected factor $3\times$ improvement in $B_1$ (see Fig.~\ref{fig:capacitance_TCAD_triangulation berdina} for schematic of the gate layout).

\subsection{Experimental Setup}
The sample was mounted on a high-frequency printed circuit board in a copper enclosure, thermally anchored to the cold finger of an Oxford Kelvinox $100$ dilution refrigerator with a base temperature $T_{\rm bath}=20$~mK. The sample was placed in the center of a wide-bore superconducting magnet, oriented so that the $B_0$ field was applied along the [110] plane of the Si substrate, and
perpendicular to the short-circuit termination of the MW antenna. The magnet was operated in persistent mode while also feeding the nominal current through the external leads. We found that removing the supply current while in persistent mode led to a very significant magnetic field and ESR frequency drift, unacceptable given the intrinsic sharpness of the resonance lines of our
qubit. Conversely, opening the persistent mode switch led to noticeable deterioration of the spin coherence, most visible as a shortening of $T^\ast_2$ in Ramsey experiments.

Room-temperature voltage noise was filtered using an anti-inductively wound coil of thin copper wire with a core of Eccosorb CRS-117 ($\sim1$~GHz cut-off), followed by two types of passive low-pass filters: $200$~Hz second-order $RC$ filters for DC biased lines, and $80$~MHz seventh-order Mini-Circuits LC filters for pulsed voltage lines (see supplementary section \ref{FreqResp} for measurements on the frequency response of the SET to a square wave applied to different gates.). The filter assemblies were placed in copper enclosures, filled with copper powder, and thermally anchored to the mixing chamber. DC voltages were applied using optoisolated and battery-powered voltage sources, connected to the cold filter box via twisted-pair wires. Voltage pulses were applied using an arbitrary waveform generator (LeCroy ArbStudio $1104$), connected to the filter box via semi-rigid coaxial lines. ESR excitations were generated using an Agilent E8267D analog signal generator, and NMR excitations were produced by an Agilent MXG N5182A vector signal generator. Both excitation signals were combined using a power-combiner and fed to the MW antenna via a CuNi semi-rigid coaxial cable, with attenuators at the $1.5$~K stage ($10$~dB) and the $20$~mK stage (3 dB). The SET current was measured by a Femto DLPCA-200 transimpedance amplifier at room temperature, followed by a floating-input voltage post-amplifier, a sixth-order low-pass Bessel filter, and acquired using a PCI digitiser card (AlazarTech ATS9440).

\subsection{Data Acquisition Statistics}
For e$^-$ experiments the state is always initialized spin-down and all of our plots were produced by taking the spin-up proportion from $100-200$ single-shot measurement repetitions per point. For $^{31}$P experiments, plots were produced by taking the nuclear flipping probability (no initialization to a certain state) from $41$ measurement repetitions per point, and $50$ electron readouts per nuclear spin readout. See Ref.~\cite{Pla2013} for more details on nuclear spin readout and control sequences.

\subsection{\label{FreqResp}Frequency Response of the Electrostatic Gates}

All electrical gates are low-pass filtered to minimize the electron temperature. The top gate (TG), left barrier gate (LB) and the right barrier gate (RB) are filtered with $200$~Hz second-order $RC$ filters. The left donor ``slow gate'' (LDS) and the right donor ``slow gate'' (RDS) are filtered with nominally $10$~kHz second-order $RC$ filters. The left donor ``fast gate'' (LDF), the right donor ``fast gate'' (RDF), and the SET tuning gate (TGAC) are filtered with nominally $80$~MHz seventh-order Mini-Circuits LC filters (see Fig.~\ref{fig:capacitance_TCAD_triangulation berdina} for schematic of the gate layout).
\begin{suppfig}[!b]
\begin{center}
\includegraphics[width=1\textwidth]{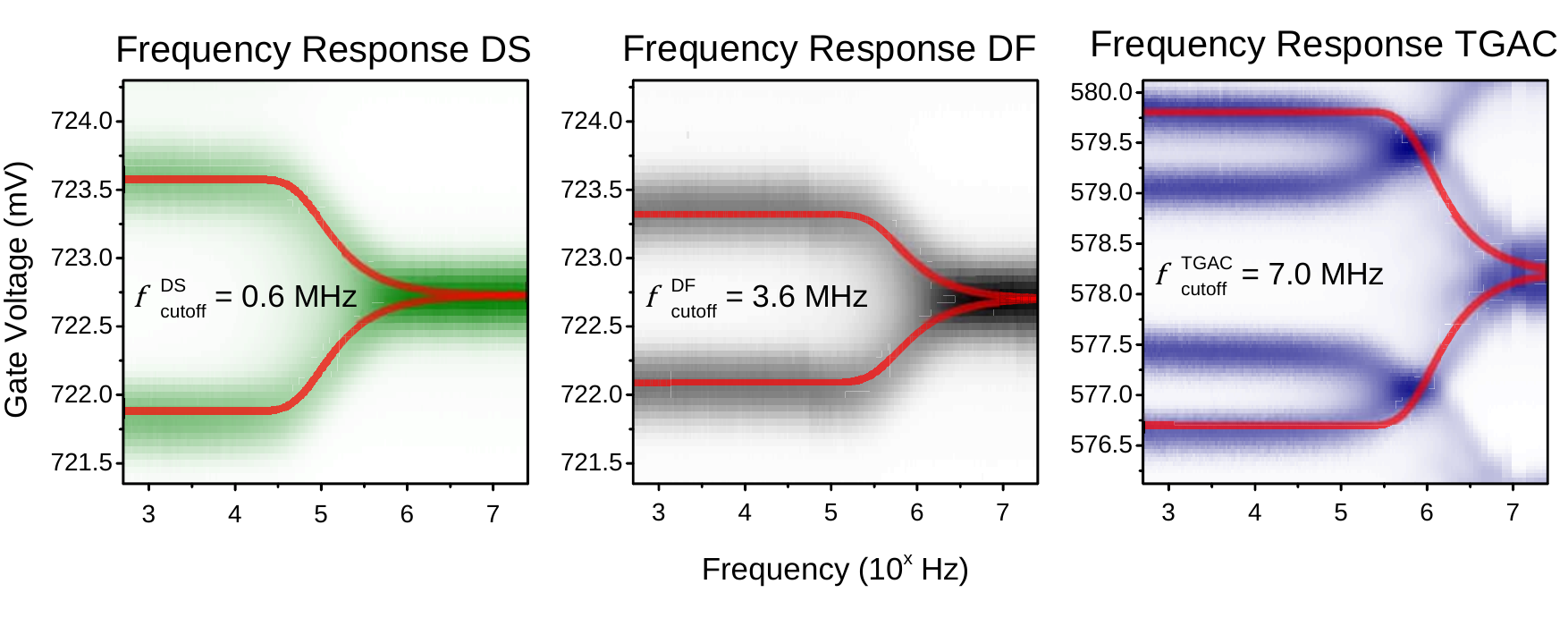}
\caption{\label{figFreqResp} \textbf{Frequency reponse.}
Frequency response of the SET current to a square wave applied to donor slow (DS), donor fast (DF) and the SET tuning gate (TGAC). The red solid lines are best estimates of the cutoff frequencies.}
\end{center}
\end{suppfig}
However, the measured response of the SET to these different gates seems to differ from the engineered cutoffs. The reason for this is unknown, but could be explained by failed components or spurious conducting paths through the copper powder surrounding the lines inside the filter boxes. In Fig.~\ref{figFreqResp} we plot the current through the SET (colored regions indicate higher current) when a square wave is applied to the DS (LDS+RDS), DF (LDF+RDF) and the TGAC gates, respectively. For a frequency below the cutoff of the cable and filter, individual SET Coulomb peaks are split into doublets. Above the cutoff frequencies the square wave is strongly attenuated and the Coulomb peaks merge back into single peaks. We estimate the cutoff frequencies of the different lines by comparing the measured data with theoretically modelled second-order $RC$ filters, where $R_1 = 20$~k$\Omega$, $C_1 = 1$~pF, $R_2 = 20$~k$\Omega$, and $C_2 = 1/(R_2 f_{\rm cutoff})$. We find $f^{\rm DS}_{\rm cutoff} = 0.6$~MHz, $f^{\rm DF}_{\rm cutoff} = 3.6$~MHz, and $f^{\rm TGAC}_{\rm cutoff} = 7.0$~MHz, and plot the corresponding frequency responses as red lines in Fig.~\ref{figFreqResp}.

For the electrically-controlled measurements in the main text, a combination of DS, DF, and TGAC was used as ``$A$-gate''. For the measurements in Figs.~\ref{figure02} \& \ref{figureESR} the voltages applied to the gates were $V_{\rm DS}=V_A$, $V_{\rm DF}=V_A$, and $V_{\rm TGAC}=0.8 V_A$. For the measurements in Figs.~\ref{figure03} \& \ref{figure04}, and all other measurements in the supplementary information we reduced the voltage applied to the DS gate to improve the frequency response of the system. The applied voltages for these measurements were $V_{\rm DS}=0.5 V_A$, $V_{\rm DF}=V_A$, and $V_{\rm TGAC}=0.8 V_A$.

\newpage
\subsection{\label{Triag}Triangulation of Donor Position}

We can triangulate the position of the ion-implanted donor based on the techniques and methods introduced in Ref.~\cite{Mohiyaddin2013}. The triangulation is obtained by combining two different techniques, each one predicting a locus or several loci of donor locations compatible with a measurable physical property of the system. We use a classical, finite-element electrostatic simulation software (TCAD)~\cite{tcad_manual_2004}, to model the electrostatic potentials in order to match the spin readout criterion (ground state energy of the donor-bound electron aligned with the Fermi level of the SET island), and a geometric capacitance extraction method (FASTCAP)~\cite{Nabors_IEEE_TCADICS_1991}, to match the measured capacitive coupling between the donor and the surrounding gate electrodes.

For this device, we experimentally observed that the donor was more strongly capacitively coupled to TGAC than to the donor tuning gates LDF, LDF, RDS, and RDF. This indicates that the donor is positioned on the TGAC-side of the SET island. This may be due to misalignment of the implant window during fabrication, or to a donor belonging to the background doping of the epilayer. Matching the relative donor capacitances, obtained from charge stability experiments (see Table~\ref{tabdonorC}) with FASTCAP simulations, we obtain 4 loci for the possible donor location (see Figure \ref{fig:capacitance_TCAD_triangulation berdina}).

\begin{suppfig}[t!]
\begin{center}
\includegraphics[width=\textwidth, keepaspectratio = true]{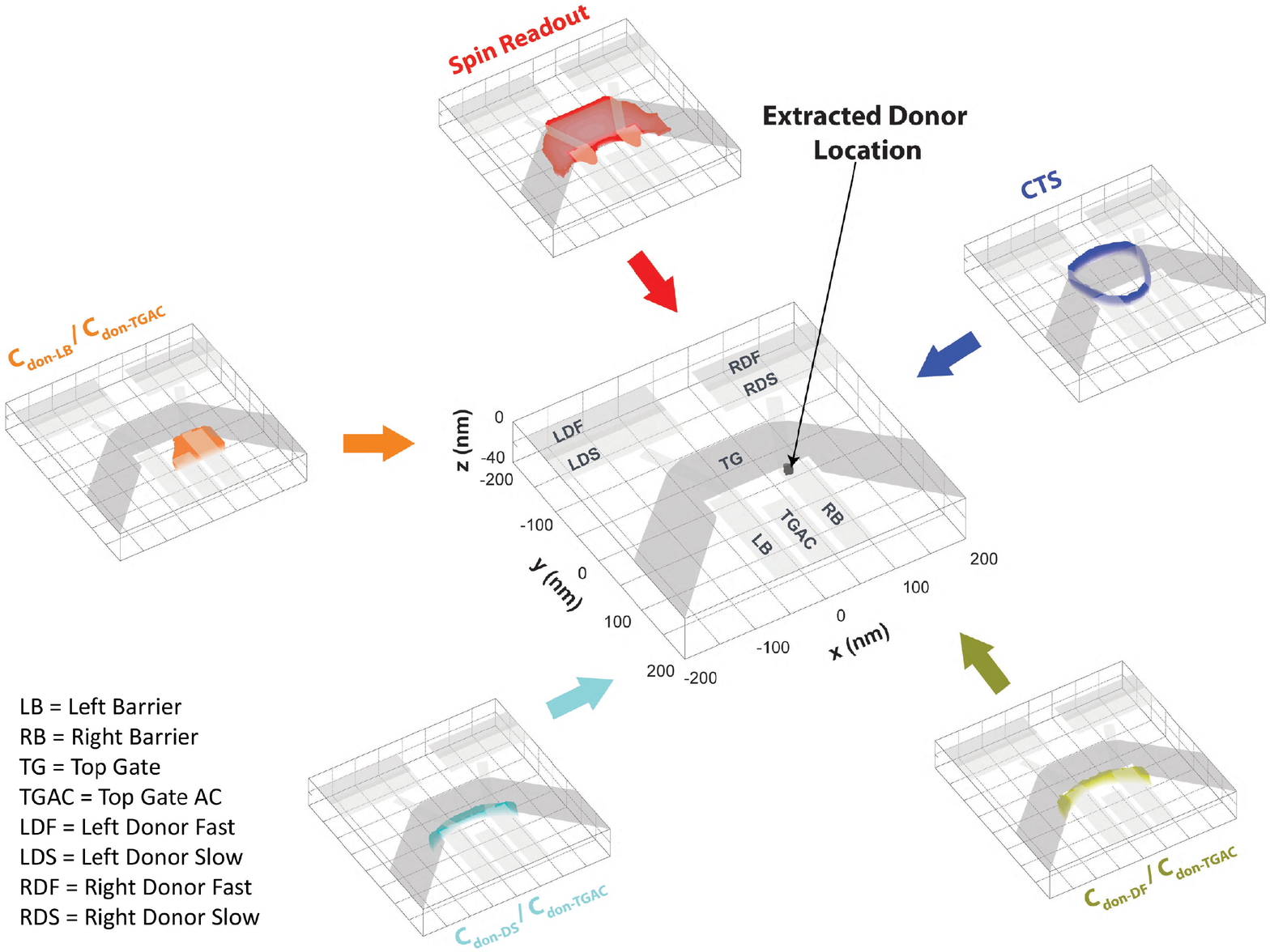}
\caption{\textbf{Donor triangulation.} Donor position extracted from a combination of donor ground state energy and relative donor capacitances to various gates. The combination of donor capacitances and spin readout criterion aids to reduce the uncertainty of the donor position to $\pm$4, $\pm$2.5 and $\pm$3.5 nm in the three cartesian axis x, y and z. }
\label{fig:capacitance_TCAD_triangulation berdina}
\end{center}
\end{suppfig}

A 5th locus can be obtained from matching the spin readout criterion. The device had a threshold voltage of $\sim 1.3$~V. To faithfully describe the electrostatics of the experimental device, we include a negative interface charge density of $Q_{\mathrm{ox}} =\mathrm{ -1.8 \times 10^{12}}$~cm$^{-2}$ in the TCAD model, necessary to match the threshold voltage \cite{Mohiyaddin2013}. This charge density is consistent with estimates from deep level transient spectroscopic measurements \cite{Johnson2010}.
In a gated nanostructure, the donor can be susceptible to strain which can modify the conduction band energy\cite{Thorbeck2014,Lo2014b}. Therefore, for our metrology, we choose a fairly large error bar for the spin readout criterion, accepting locations with $E_c = 45.6 \pm 20$~meV. The locus that matches the spin readout criterion is plotted in Fig.~\ref{fig:capacitance_TCAD_triangulation berdina}. We will see in Section~\ref{TCAD} that the electric field at our final donor location is $\sim6$~MV/m, resulting in a slope of the conduction band of $\sim6$~meV/nm.  Hence, our large error bar in the spin readout criterion would only translate to a small error in donor position of $\sim3$~nm.

 \begin{table*}
 \begin{center}
 \begin{tabular}{|c|c|}
 \hline
 Gates & Relative Capacitance\\ \hline \hline
 $C_{donor - \rm LB}/C_{donor - \rm TGAC}$ 		                        & $0.65 \pm$ 0.25  \\ \hline
 $(C_{donor - \rm LDS} + C_{donor - \rm RDS})/C_{donor - \rm TGAC}$		& $1.53 \pm$ 0.25  \\ \hline
 $(C_{donor - \rm LDF} + C_{donor - \rm RDF} )/C_{donor - \rm TGAC}$	& $0.36 \pm$ 0.075 \\ \hline
 $CTS = C_{donor - island}/C_{donor}$	                                & $0.36 \pm$ 0.06  \\ \hline
 \end{tabular}
 \end{center}
  \caption{Relative gate capacitances used for triangulation of the donor position.}\label{tabdonorC}
 \end{table*}

The 5 shells in Fig.~\ref{fig:capacitance_TCAD_triangulation berdina} intersect to within a region [50 $\pm$ 4, -31.5 $\pm$ 2.5, -8.5 $\pm$ 3.5], and represent the possible set of donor locations in the device. The donor is located under the right barrier gate RB, towards the TGAC gate.

\newpage
\subsection{\label{TCAD}Electric Field Simulations}

We can use the triangulated donor positions and our TCAD model of the Si/SiO$_2$ structure with the Al-gates (see section~\ref{Triag}) to estimate the local electric field at the donor site. Fig.~\ref{figEF_V2} shows the magnitude and direction of the calculated electric field for $V_A=50$~mV (additional to $V_{\rm LB} = 0.92$~V, $V_{\rm RB} =	0.92$~V, $V_{\rm TG} = 1.798$~V, $V_{\rm DS} =	0.579$~V, $V_{\rm DF} = 0.575$~V, $V_{\rm TGAC} = 0.45$~V, compare also with section~\ref{FreqResp}) inside a coordinate range that comprises the triangulated donor positions.
\begin{suppfig}[!b]
\begin{center}
\includegraphics[width=0.85\textwidth]{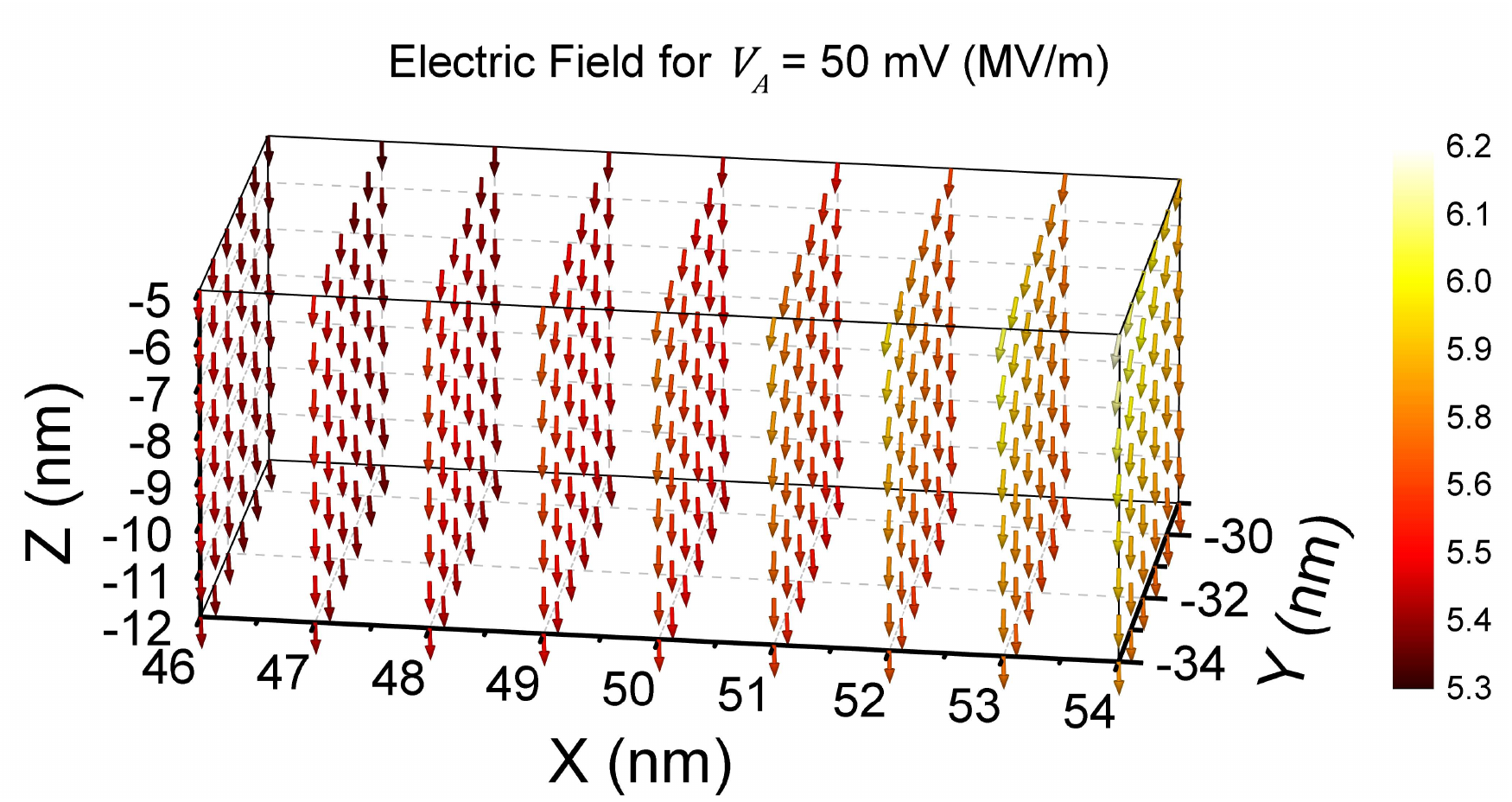}
\caption{\label{figEF_V2} \textbf{Electric field simulations.}
Calculated electric field for $V_A=50$~mV.}
\end{center}
\end{suppfig}
Since the donor is located beneath the barrier gate, the positive bias applied to this gate ($V_{\rm RB}=0.92$~V) causes the electric field to point in negative $z$-direction, and the electron wavefunction is pulled towards the interface and away from the nucleus. This explains the very low initial hyperfine coupling $A=96.9$~MHz, which is significantly different from the bulk value of $117.53$~MHz. Increasing $V_{\rm RB}$ should pull the electron even further away from the nucleus reducing $A$ even more. This means that we are starting in a situation slightly different from what is schematically depicted in Fig.~\ref{figure01}B, as the electron is already displaced towards the interface from the beginning.
\begin{suppfig}[!t]
\begin{center}
\includegraphics[width=0.85\textwidth]{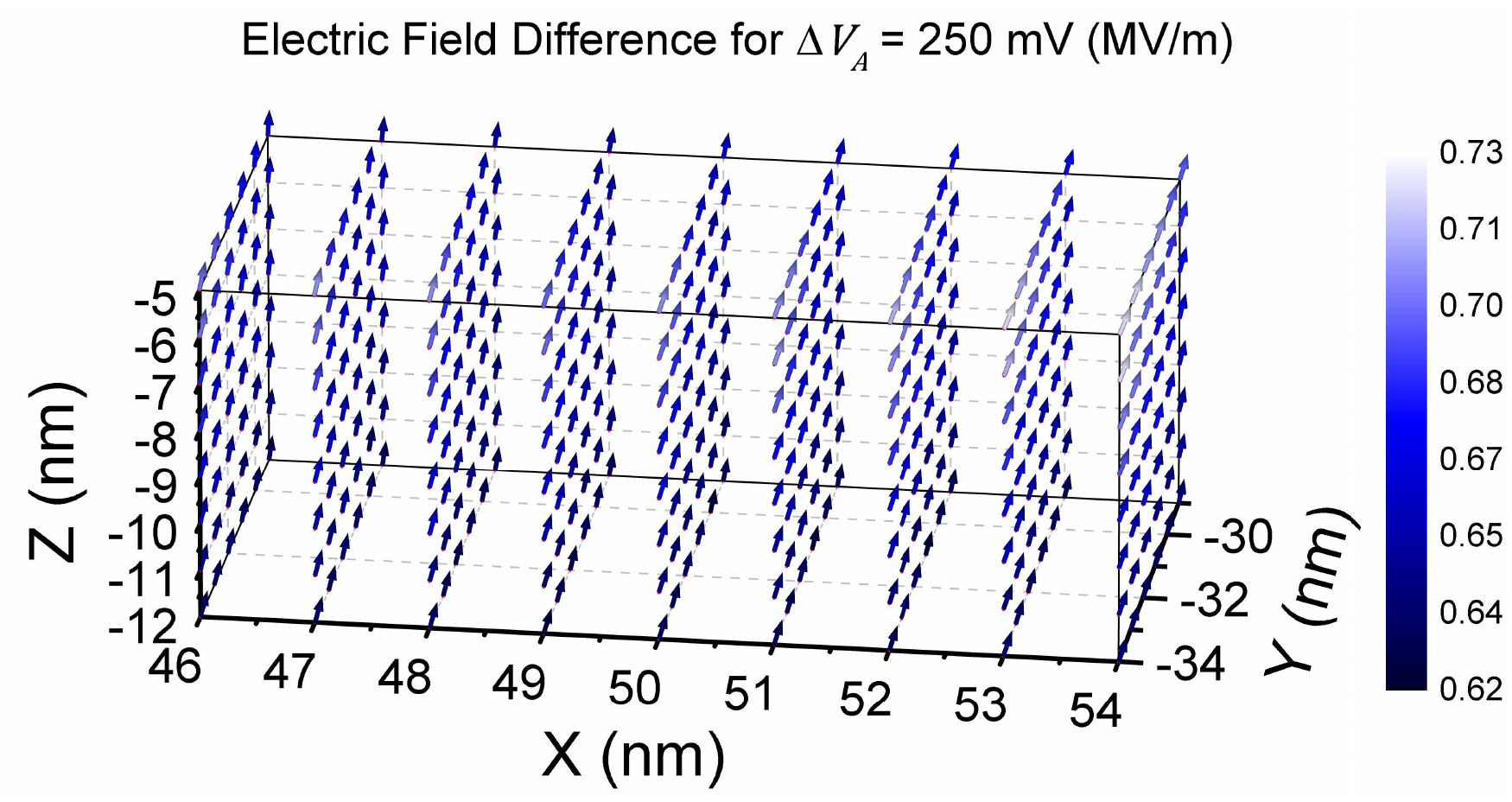}
\caption{\label{figEF_V7-V2} \textbf{Electric field simulations.}
Calculated electric field difference for $\Delta E = E_{V_A = 300\rm mV} - E_{V_A = 50\rm mV}$.}
\end{center}
\end{suppfig}

It may seem contradictory that, in experiment, we have measured a positive tuning parameter $\alpha_{A} = 0.91\pm0.07$~MHz/V, meaning that a more positive $V_A$ leads to an increase in the hyperfine coupling. This is explained by the fact that the $A$-gate consists of the DS-gates, DF-gates, and the TGAC-gate (see section~\ref{FreqResp}), which are all located some distance away from the donor. In this case an increase in $V_A$ will effectively compensate the electric field under the right barrier. We confirm this in Fig.~\ref{figEF_V7-V2}, where we plot the change in electric field when $V_A$ is increased from $V_A=0.50$~mV to $V_A=300$~mV, i.e. $\Delta V_A=250$~mV. The change in electric field is clearly positive, reducing the magnitude of the vertical electric field when $V_A$ is increased, and therefore pushing the electron back towards the donor, resulting in a positive value of $\alpha_A$.

\begin{suppfig}[!t]
\begin{center}
\includegraphics[width=0.7\textwidth]{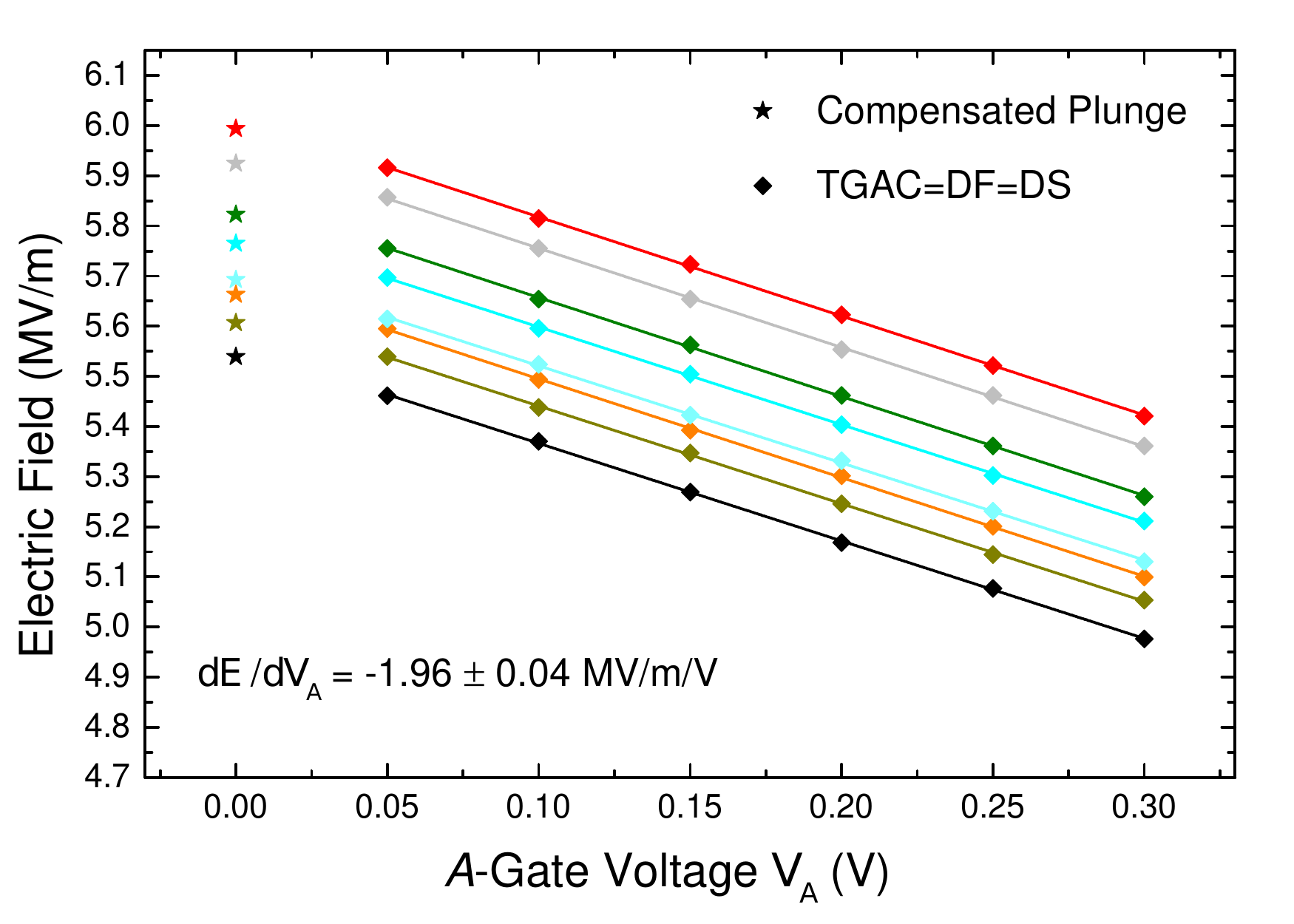}
\caption{\label{figVAWG-EF} \textbf{Electric field simulations.}
Calculated electric field as a function of $V_A$ for possible donor locations. The change in electric field is very similar for all donor locations.}
\end{center}
\end{suppfig}

In Fig.~\ref{figVAWG-EF} we plot the electric field $E$ at possible donor locations as a function of $V_A$. The stars correspond to the compensated plunge position (``Pulse ESR/NMR'' position in Fig.~\ref{figure02}A), that keeps the potential of the SET island constant with respect to the Fermi level of source and drain~\cite{Morello2010,Pla2012}. The diamonds correspond to the electric fields calculated for the different values of $V_A$. The absolute value of $E$ varies significantly between the different donor positions, and converting $V_A$ into $E$ would be subject to a large error. However, the tunability of $E$ is very similar for all locations. Therefore, we fit this set of simulations to extract an average value for the tunability $dE/dV_A = -2.62(5)$~MVm$^{-1}$/V, where the error is the standard deviation of the slopes of individual fits. This value is used to convert from $V_A$ to $dE$ with good accuracy, and to calculate the ``Electric Field Change''-axis of Fig.~\ref{figure02}B.

\newpage
\subsection{\label{HFsim}Atomistic Simulations of the Hyperfine Coupling}

The spin Hamiltonian of a $^{31}$P donor electron spin $\textbf{S}$  and nuclear spin $\textbf{I}$ in an electrostatic potential $\phi$ and magnetic field $\textbf{B}_0$ is given by:

\begin{equation}
H_{\mathrm{P-spin}} = \gamma_e(\phi)\textbf{S}\cdot\textbf{B}_0 - \gamma_n\textbf{I}\cdot\textbf{B}_0 + A(\phi)\textbf{I}\cdot\textbf{S}
\label{Hamiltonian_donor_spin}
\end{equation}

The first and second terms in Equation \ref{Hamiltonian_donor_spin} are the electronic and nuclear Zeeman terms, the third term is the contact hyperfine interaction between the two spins, and $\gamma_e$ and $\gamma_n$ are the electron and nuclear gyromagnetic ratios, respectively. The contact hyperfine coupling \cite{Hale1969} is expressed as $A(\phi) = \frac{8\pi}{3}\gamma_e(\phi)\gamma_n\lvert{\psi(r_0,\phi)\rvert}^2$, where $\lvert{\psi(r_0,\phi)\rvert}^2$ is the probability density of the electron wave function evaluated at the donor site $r_0$. 

The relative tunability of the gyromagnetic ratio $\alpha_{\gamma_e} / \gamma_e(0)$ is two orders of magnitude smaller than the relative tunability of the hyperfine coupling $\alpha_A / A(0)$ (refer to Fig.~\ref{figure02} and Ref.~\cite{Rahman2009}). The hyperfine coupling relative to the bulk value ($A(0) = 117.6$ MHz) can, therefore, be approximated as \cite{Mohiyaddin2013,Martinis2004,Rahman2007}
\begin{equation}
\frac{A(\phi)}{A(0)} = \frac{{\lvert\psi(r_0,\phi)\rvert}^2} {{\lvert\psi(r_0,0)\rvert}^2}.
\label{scaling}
\end{equation}

We use a numeric implementation of tight binding - packaged as a software tool called NanoElectronic MOdeling-3D (NEMO-3D)~\cite{Klimeck2007, Klimeck2007b} - to calculate the hyperfine coupling of the donor at the location triangulated in Section~\ref{Triag} and for the electric fields simulated in Section~\ref{TCAD}. Typical simulation domains of 30 nm $\times$ 30 nm $\times$ 30 nm consisting of approximately 1.4 million atoms were considered. Each NEMO simulation with the above domain takes $\sim$ 2 hours, when run on a computing cluster with 48 processors.
\begin{suppfig}[b!]
\begin{center}
\includegraphics[width=0.6667\columnwidth]{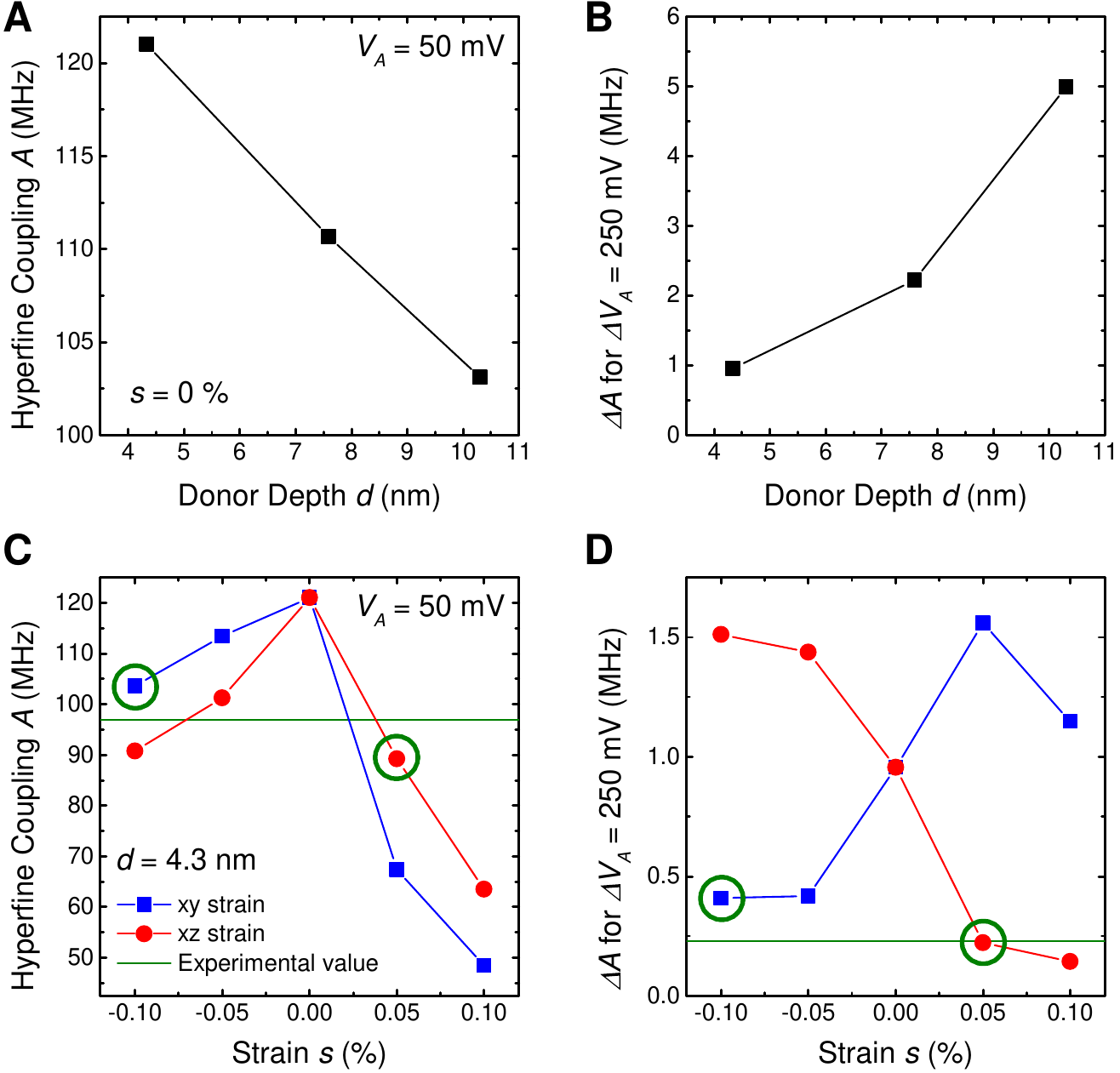}
\caption{\textbf{Atomistic tight binding simulations of the hyperfine coupling for a donor at the location determined in Section~\ref{Triag}, for the electric fields simulated in Section~\ref{TCAD}, and subject to lattice strain.}
\textbf{A, B} Hyperfine coupling and tunability of the hyperfine coupling as a function of donor depth under the SiO$_2$ in the unstrained lattice.
\textbf{C, D} Hyperfine coupling and tunability of the hyperfine coupling as a function of lattice strain for a donor depth $d=4.3$~nm under the SiO$_2$. The green lines indicate the experimental values and the green circles highlight simulations in good agreement with the experimental results.}
\label{figHF}
\end{center}
\end{suppfig}

In Fig.~\ref{figHF}A we plot the calculated hyperfine coupling for $V_A = 50$~mV for three different donor depths $d$ at the triangulated position. Confinement by the interface enhances  ${\lvert\psi(r_0,\epsilon)\rvert}^2$ beyond the bulk value for near shallow donors ($d=4.3$~nm), resulting in a hyperfine coupling greater than $117.6$~MHz. However, for donors further away from the interface, the electric field pulls the electron away from the nucleus, resulting in a decrease in hyperfine coupling~\cite{Rahman2007,Mohiyaddin2014}. The tunability of the hyperfine coupling $\Delta A$, i.e. how strongly $A$ is modified when the electric field is changed, scales inversely proportional with the confinement of the electron. We plot $\Delta A = A_{V_A = 300\rm mV} - A_{V_A = 50\rm mV}$ for three different donor depths in Fig.~\ref{figHF}B. We notice that the calculated tunability of $A$ is much larger than the experimentally measured one (compare Fig.~\ref{figure02}). We believe this to be due to strain of the Si lattice, which also influences the hyperfine coupling by distorting the donor wave function~\cite{Dreher2011}.

We cannot quantify the strain in our device, but Thorbeck \textit{et al.}~\cite{Thorbeck2014} estimate that strain due to metal surface gates in nanoelectronic devices can be as much as $\sim 0.1$~\%. Especially with the donor located directly under a barrier gate (see supplementary section~\ref{Triag}) we expect significant strain at the donor location. We repeat the hyperfine calculations for the donor $d=4.3$~nm below the interface for different values of homogeneous strain. Here, compressive strain ($s < 0$) in the xy-plane will lead to tensile strain ($s > 0$) in the z-direction. We plot the calculated hyperfine coupling for $-0.1\% < s < 0.1\%$ for strain applied to the xy-plane (blue squares) and the xz-plane (red circles) in Fig.~\ref{figHF}C ($V_A=50$~mV). Any type of strain decreases the wavefunction overlap of electron and nucleus and reduces the hyperfine coupling $A$. In Fig.~\ref{figHF}D, we plot the corresponding tunability of the hyperfine coupling $\Delta A$ (for $\Delta V_A = 250$~mV). For certain values of $s$ both the absolute value of $A$ and the tunability $\Delta A$ are in good agreement with the experimental values of $A_{exp} = 96.9$~MHz and $\Delta A_{exp} = 0.228$~MHz (highlighted with green circles). While the presence of strain in the device seems to be able to match the simulated values with the experimental values, we cannot conclusively attribute the observed effects to a specific value of strain. Further and more detailed simulations would be required to untangle the effects of electric field, strain and the SiO$_2$ interface on the hyperfine coupling.

Electric fields, strain and the SiO$_2$ interface have profound consequences for future multi-qubit devices as they can lead to a strong variability in the hyperfine coupling $A$ and, therefore, to distinctively different qubit resonance frequencies. It can, however, be expected that multi-qubit devices with deterministically positioned donors will have a smaller variability in the hyperfine coupling as the donors can be placed at locations with little or at least similar strain. Furthermore, a gate layout optimized for maximizing the Stark shift should give a tunability of the resonance frequencies of a few MHz and be sufficient for the operation of a multi-qubit quantum computer with a monochromatic global microwave field. In a scenario where the variability is too large or the tuning range too small, the global microwave could be operated as multi-tone continuous-wave driving field at regular frequency spacing.

\newpage
\subsection{\label{TES-ESR}Electron Spin Resonance Spectrum \& Rabi Oscillations - Experiment \& Theory}

In Fig.~\ref{figureESR}B we show a gate-controlled measurement of the ESR spectrum, obtained by shifting $\nu_{e}$ with $V_A$. The nuclear spin is in the $|{\Uparrow\rangle}$ state for the duration of the experiment, and we apply a continuous-wave MW driving field at frequency $\nu_{\rm MW} = \nu_{e2}^{\rm ref}-200$~kHz. Here, $\nu_{e2}^{\rm ref}$ is the ESR frequency for $|{\Uparrow\rangle}$ obtained from a conventional pulsed Ramsey experiment at the compensated plunge position (``Pulse ESR/NMR'' position in Fig.~\ref{figure02}A). We start the sequence (see Fig.~\ref{figureESR}A) at $V_A = 0$~V to load an electron in the $|{\downarrow\rangle}$ state by spin-dependent tunneling~\cite{Morello2010} (``Read/Init.'' position in Fig.~\ref{figure02}A). We then apply a positive $V_{A}$ pulse to shift $\nu_{e2}(V_A)$ towards $\nu_{\rm MW}$ for the duration $t_p$, which results in a coherent manipulation of the spin. We then pulse $V_A$ back to $0$~V to perform single-shot readout of the electron spin~\cite{Morello2010}, and we repeat the whole sequence 200 times to extract the electron spin-up fraction $P_{\uparrow}(V_A)$. For $V_{A} = V_r \approx 155$~mV, an increased count of spin $|{\uparrow\rangle}$ electrons indicates that $\nu_{e2}(V_A)$ becomes resonant with the MW source. By optimizing the duration of the gate pulse we can ensure that the electron spin undergoes a $\pi$-rotation while on resonance, yielding $P_{\uparrow} \approx 1$.

The maximum $P_{\uparrow}$ occurs at a frequency shift $\Delta\nu_e = -217$~kHz instead of the expected $-200$~kHz. This shift and the presence of side lobes at larger $V_A$, is caused by the finite time response of the electrical control gates, which are low-pass filtered to minimize electron heating (see supplementary section \ref{FreqResp} for details). When measuring the gate-controlled ESR spectrum beyond $\nu_{\rm MW}$ ($V_A > V_r$), we sweep $\nu_{e1}(V_A)$ through the resonance and back again. The limited bandwidth of the gates causes slow crossings through the resonance condition, resulting effectively in a Landau-Zener-St\"uckelberg interferometry experiment~\cite{Shevchenko2010}. Time-evolution simulations of the whole sequence (red line in Fig.~\ref{figureESR}B), taking into account the bandwidths of the different gates, show excellent agreement with the measured data. In a setup optimized for electrical control, the rise-time of the $A$-gate should be chosen much shorter than the qubit Rabi period while on resonance.

\begin{suppfig}[!t]
\begin{center}
\includegraphics[width=0.5\textwidth]{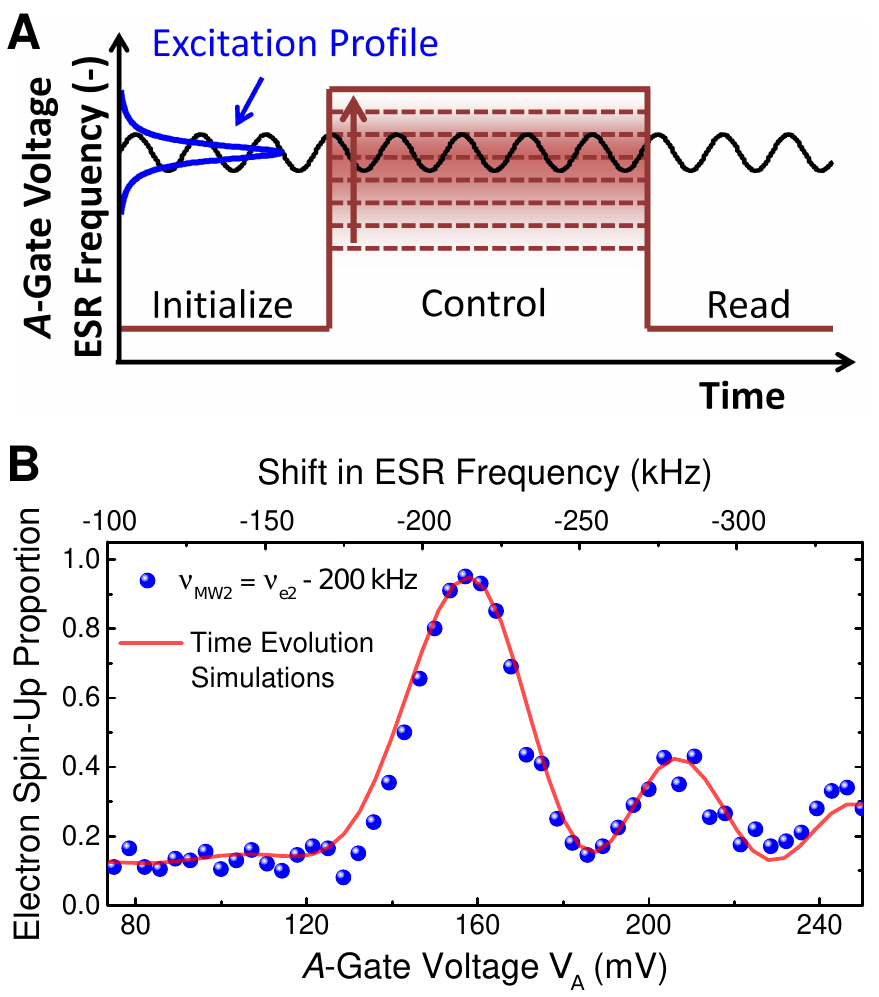}
\caption{\label{figureESR} \textbf{Electrically-controlled ESR spectrum.}
\textbf{A}, Schematic of the sequence used to measure the electrically-controlled ESR spectrum.
\textbf{B}, ESR spectrum obtained by electrically tuning the ESR transition (for $|{\Uparrow\rangle}$) in resonance with the continuous-wave MW source ($\nu_{\rm MW} = \nu_{e2} -200$~kHz) with a short voltage pulse. Time evolution simulations (red line) are in excellent agreement with the data.}
\end{center}
\end{suppfig}

In the following paragraphs we present the theoretical model that we have set up to describe the temporal response of the donor system to a voltage pulse on the $A$-gate, and the temporal evolution of spin qubits when they are tuned into resonance with a CW magnetic driving field.

We model the experimental data using the density matrix formalism with the Hamiltonian
\begin{equation}
\label{hami}
H(t) = \frac{1}{2}h \Delta\nu(t) \sigma_z + \frac{1}{2}h \Omega_0 \sigma_x,
\end{equation}
where $\Delta\nu(t) = \nu_{e2}(t)-\nu_{\rm MW}$ is the detuning between the ESR transition and the MW source, and $\Omega_0 = 23.8$~kHz is the Rabi frequency for a $B_1=0.85$~$\mu$T, which gives a $\pi$-pulse length of $21$~$\mu$s. The output power of the MW source $P_{\rm MW} = -22$~dBm was chosen very low for these experiments to reduce the power broadening of the ESR line.
The dephasing time $T_2=970$~$\mu$s of the electron spin (see Fig.~\ref{figure03}H) is included in the master equation of the Lindblad form~\cite{Kok2010}
\begin{equation}
\label{lind}
\frac{d\rho}{dt}=-\frac{i}{\hbar}[H,\rho]+\mathcal{L}(\rho),
\end{equation}
where
\begin{eqnarray}
\label{liou}
\mathcal{L}(\rho) &= &\frac{1}{2 T_2} \left(2\sigma_z \rho \sigma_z - \sigma_z \sigma_z \rho - \rho \sigma_z \sigma_z \right)\nonumber \\ &= &
-\frac{2}{T_2} \sigma_x \rho.
\end{eqnarray}
We then use the equation of motion (\ref{lind}) to numerically compute the time evolution of an electron initialized in $|{\downarrow\rangle}$.

\begin{suppfig}[!t]
\begin{center}
\includegraphics[width=0.95\textwidth]{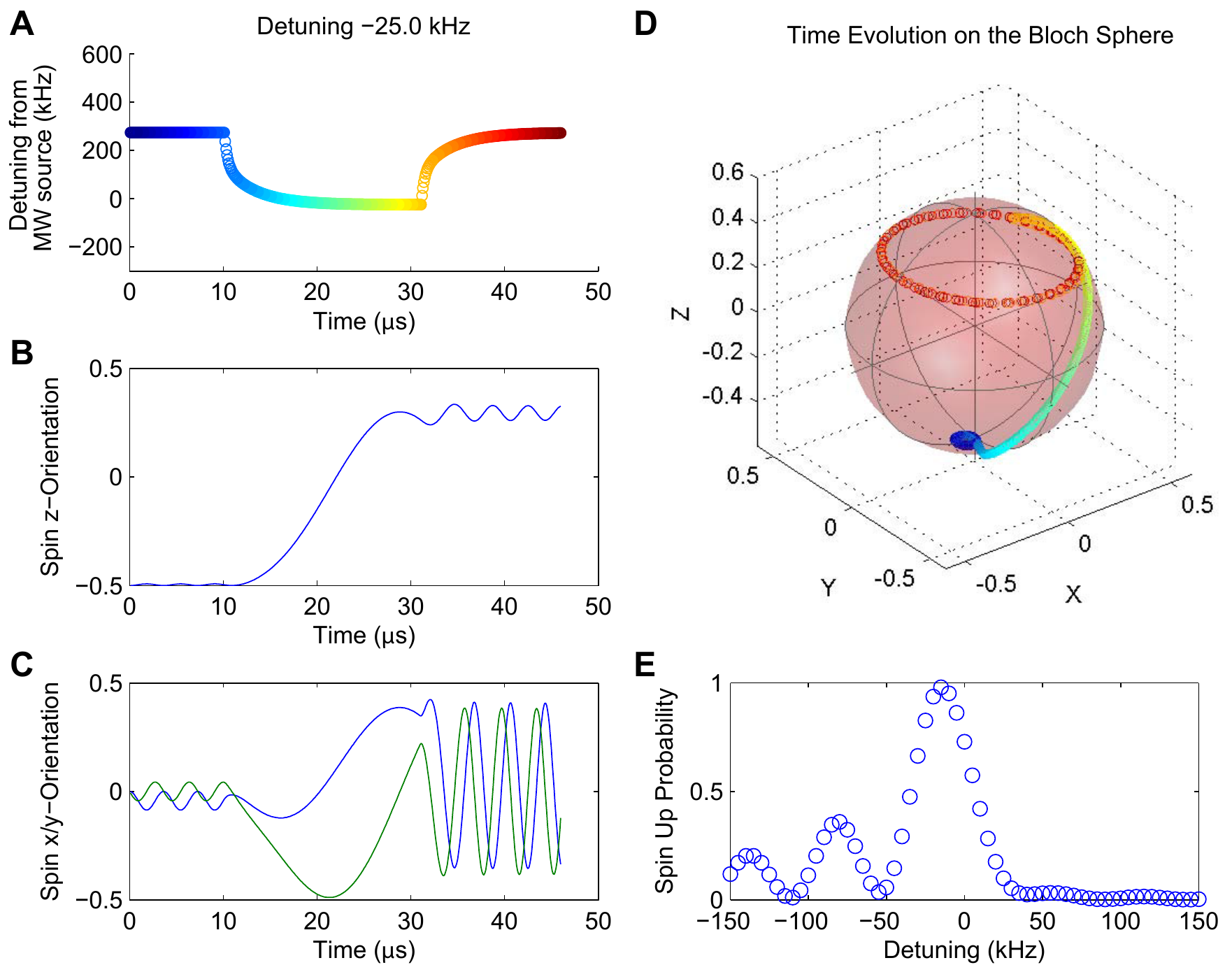}
\caption{\label{figTES-ESR} \textbf{Time evolution simulations of the electrically controlled ESR spectrum.}
Simulations depicting the time evolution of the electron spin around the Bloch sphere for a detuning pulse of $t_{pulse}=21$~$\mu$s and amplitude $249$~kHz that tunes the ESR transition to $25$~kHz lower frequency than the MW drive.}
\end{center}
\end{suppfig}
From the measurements in Fig.~\ref{figure02}B, we know $d\nu_{e2} / d V_A$ and can allocate the total shift in $\nu_{e2}$ to DS, DF, and TGAC in accordance to their respective relative capacitive couplings to the donor (see section~\ref{Triag}) and temporally filtered to their respective cutoff frequencies (see section~\ref{FreqResp}). This allows us to generate the detuning trace $\Delta\nu(t)$ for any measurement and detuning sequence that we want to model:
\begin{eqnarray}
\label{freqtrace}
\Delta\nu(t) &= &\frac{(d\nu_{e2} / d V_A)}{C_{donor - \rm DS} /C_{donor - \rm TGAC}+C_{donor - \rm DF} /C_{donor - \rm TGAC}+1}\\
& &\left((C_{donor - \rm DS} /C_{donor - \rm TGAC}) V^{\rm DS}_A(t) + (C_{donor - \rm DF} /C_{donor - \rm TGAC}) V^{\rm DF}_A(t) + V^{\rm TGAC}_A(t)\right) \nonumber \\
&= & \frac{-1.36{\rm~MHz/V}}{1.53+0.36+1} \left(1.53 V^{\rm DS}_A(t) + 0.36 V^{\rm DF}_A(t) + V^{\rm TGAC}_A(t)\right)\nonumber.
\end{eqnarray}
Here, $V^{\rm DS}_A(t)$, $V^{\rm DF}_A(t)$, and $V^{\rm TGAC}_A(t)$ are the low-pass filtered ($f^{\rm DS}_{\rm cutoff} = 0.6$~MHz, $f^{\rm DF}_{\rm cutoff} = 3.6$~MHz, and $f^{\rm TGAC}_{\rm cutoff} = 7.0$~MHz) voltage traces that are applied to the gates.

Fig.~\ref{figTES-ESR}A shows $\Delta\nu(t)$ for an electrically-controlled ESR measurement. The time evolution simulation starts with an electron in the $|{\downarrow\rangle}$ state at $\Delta\nu=274$~kHz for the first $10$~$\mu$s. $\Delta\nu$ is then changed to $-25$~kHz (in this specific example) for a time of $t_{pulse}=21$~$\mu$s ($\pi$-pulse) before it is tuned back to $274$~kHz for $16$~$\mu$s. Fig.~\ref{figTES-ESR}B,C,D show the time
\begin{suppfig}[!b]
\begin{center}
\includegraphics[width=1\textwidth]{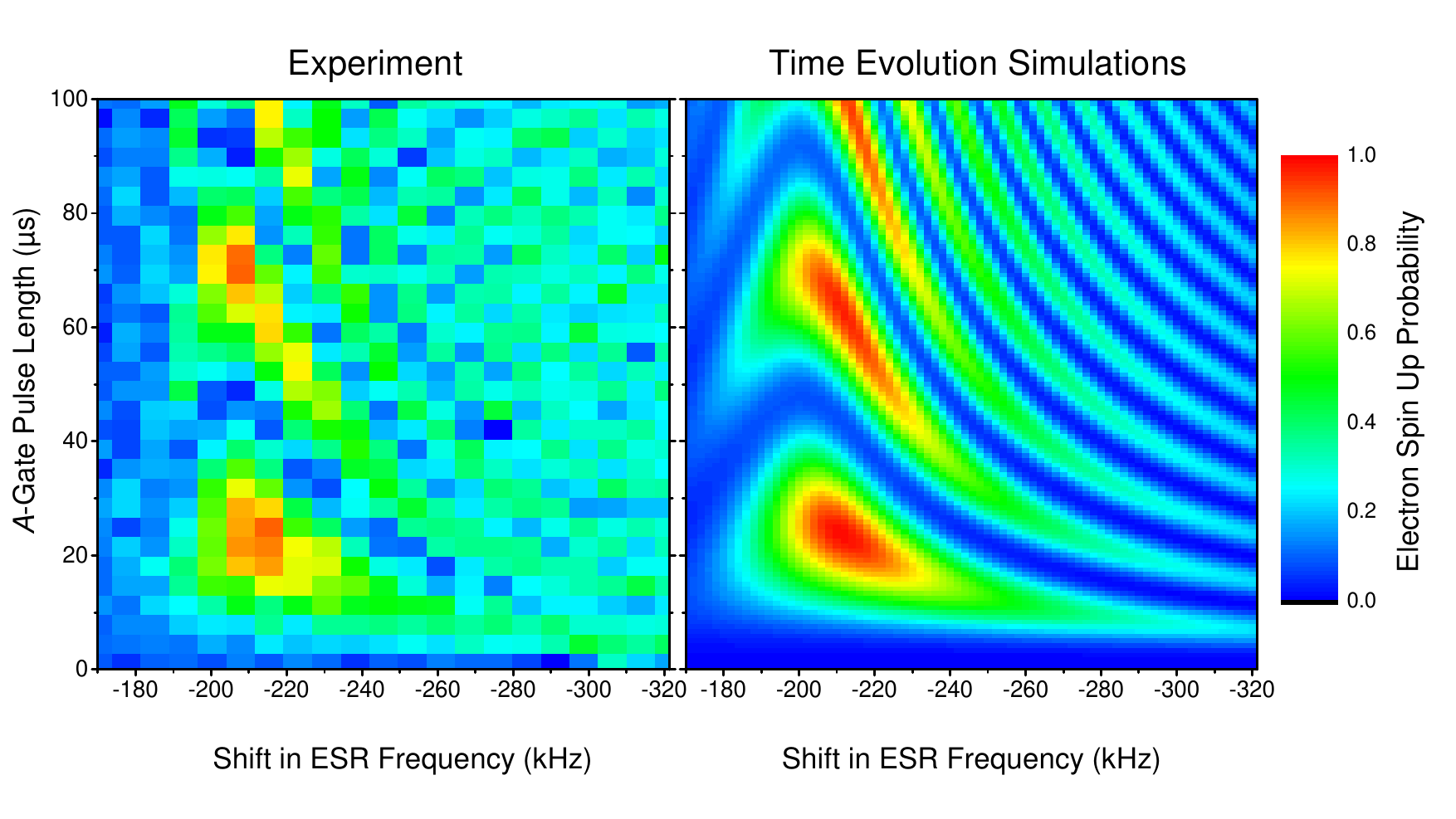}
\caption{\label{figECRabi2D} \textbf{Electrically-controlled Rabi spectrum.}
Experimental data and time evolution simulations on the electrically controlled Rabi spectrum.}
\end{center}
\end{suppfig}
evolution of the electron spin during this sequence. The final z-orientation of the electron spin is then plotted in Fig.~\ref{figTES-ESR}E, and the whole calculation is repeated for different detuning values to build up the entire ESR spectrum (comp. Fig~\ref{figTES-ESR}E and Fig.~\ref{figureESR}B). The calculated spectrum is in very good agreement with the experimental data when taking into account non-unity readout fidelity ($0.85$~$\%$) and non-zero background counts ($0.12$). The sidelobes at $\Delta\nu<0$ are well reproduced. They appear when $\nu_{e2}$ is tuned through $\nu_{\rm MW}$ at the beginning of the pulse and back at the end. The limited bandwidth of the $A$-gate causes slow crossings through the resonance condition, resulting effectively in a Landau-Zener-St\"uckelberg (LZS) interferometry experiment~\cite{Shevchenko2010}.

The measurements of Fig.~\ref{figureESR}B and the calculations of Fig.~\ref{figTES-ESR} can be also be performed as an electrically-controlled Rabi experiment, where the final electron spin orientation is measured as a function of the length of the voltage pulse. In Fig.~\ref{figECRabi2D} we plot the experimentally determined, electrically-controlled Rabi spectrum in the left panel, and the simulated one in the right panel. Again, the simulations are in excellent agreement with the experimental data and the evolution of the LZS sidelobes is well reproduced.

\newpage
\subsection{\label{Ramsey}Ramsey Experiments}

The ability to apply $V_A(t)$ sequences to the $A$-gate allows us to perform arbitrary qubit control sequences as already demonstrated in Fig.~\ref{figure03} and Fig.~\ref{figure04}. This supplies us with an alternative way to measure the frequency shift induced by $V_A$. While the measurement in Fig.~\ref{figure02}B was conducted changing $V_A$ during the control phase and pulsing the MW source, we can also conduct an electrically-controlled Ramsey experiment as introduced in Fig.~\ref{figure03}D. In Fig.~\ref{figeECRams}A we plot a series of Ramsey measurements, where $\Delta V_A = V_A^{\rm wait} - V_A^{\rm pulse}$, i.e. the detuning of the spin transition during the wait time, was changed from one measurement to the next. The frequency of the Ramsey oscillations corresponds to the detuning, and we plot the extracted values as function of $\Delta V_A$ in Fig.~\ref{figeECRams}B. The slope of the linear fit is $d\nu_{e2}/d \Delta V_A = -0.80\pm0.04$~MHz/V.

\begin{suppfig}[!b]
\begin{center}
\includegraphics[width=1\textwidth]{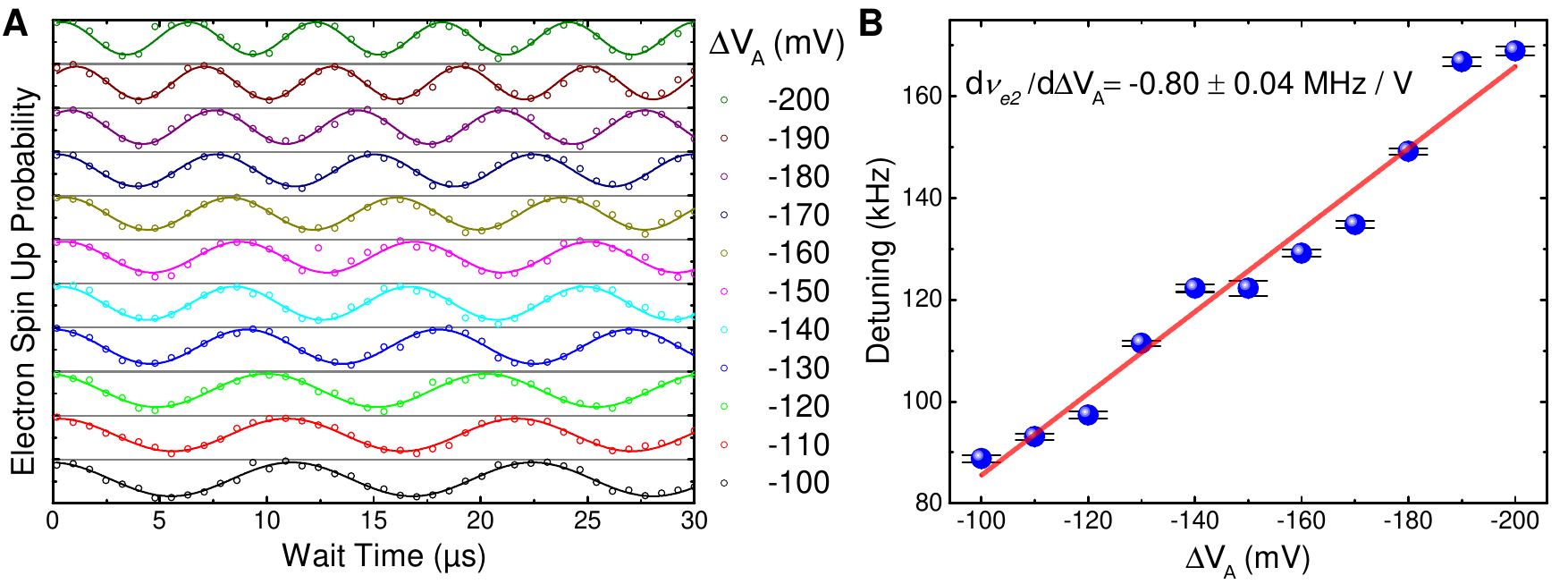}
\caption{\label{figeECRams} \textbf{Electrically controlled electron Ramseys.}
\textbf{A} Set of EC Ramsey experiments for different voltage pulses $\Delta V_{A}$ conducted on the e$^-$. \textbf{B} Corresponding frequency shift as a function of $\Delta V_{A}$.}
\end{center}
\end{suppfig}

\begin{suppfig}[!t]
\begin{center}
\includegraphics[width=1\textwidth]{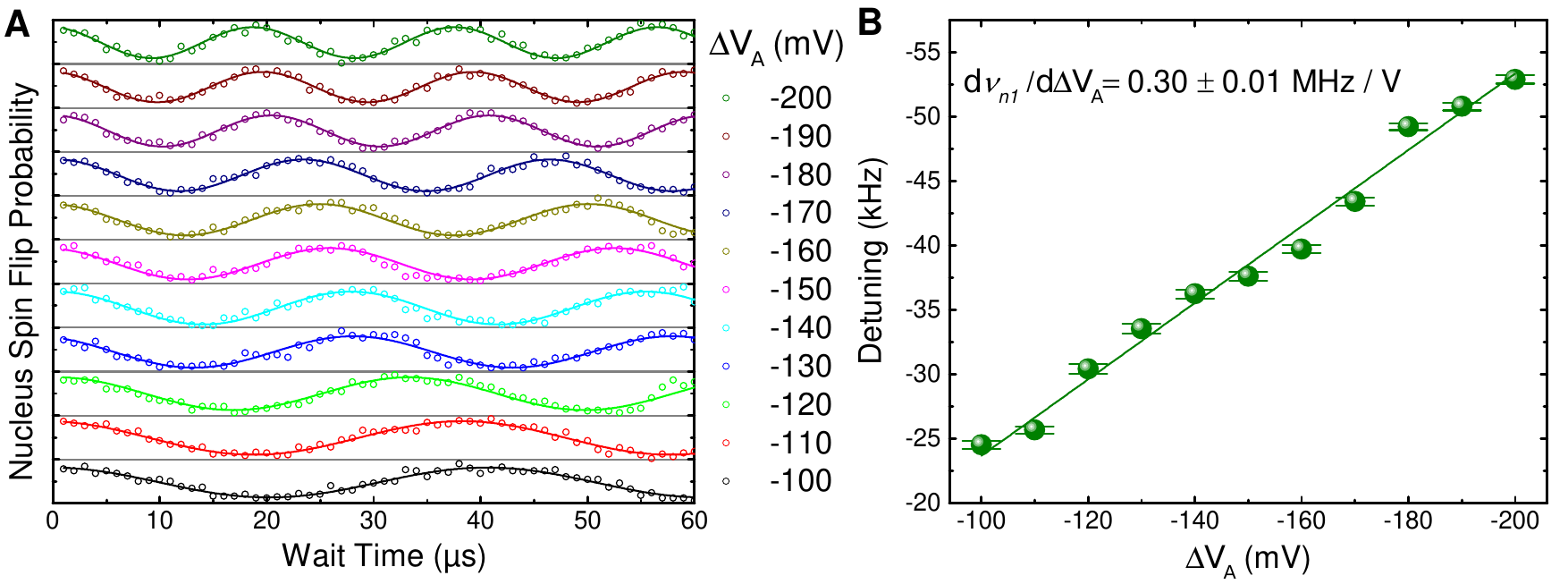}
\caption{\label{fignECRams} \textbf{Electrically controlled nuclear Ramseys.}
\textbf{A} Set of EC Ramsey experiments for different voltage pulses $\Delta V_{A}$ conducted on the $^{31}$P neutral nucleus. \textbf{B} Corresponding frequency shift as a function of $\Delta V_{A}$.}
\end{center}
\end{suppfig}

Fig.~\ref{fignECRams} shows a similar data set for the $^{31}$P neutral nucleus. The slope of the linear fit is $d\nu_{n1}/d \Delta V_A = 0.30\pm0.01$~MHz/V, which allows us to calculate the tuning parameters
$\alpha_{A} = dA/d \Delta V_A = 2d\nu_{n1}/d \Delta V_A = 0.60\pm0.02$~MHz/V, and
$\alpha_{\gamma_e} B_0 = d\gamma_e B_0/d \Delta V_A= d\nu_{e2}/d \Delta V_A - d\nu_{n1}/d \Delta V_A = -1.10\pm0.05$~MHz/V, with
$\alpha_{\gamma_e} = d\gamma_e/d \Delta V_A = -0.71\pm0.04$~MHz/V/T at $B_0 = 1.55$~T.
All the values extracted in this section are slightly smaller than the values extracted from Fig.~\ref{figure02}. This is because we reduced the voltage applied to the DS gate ($V_{\rm DS}=0.5 V_A$ compared to $V_{\rm DS}=V_A$ for the measurements in Fig.~\ref{figure02}) to improve the frequency response of the system for all electrically-controlled measurements involving pulsing sequences (see also section~\ref{FreqResp}).

\newpage
\subsection{\label{DD}Coherence Times Measurements}

In addition to the Hahn echo and CPMG measurements introduced and presented in Fig.~\ref{figure03}G,J,K, we performed Ramsey experiments to extract $T_2^\ast$ and dynamical decoupling experiments with a different number of refocussing pulses. The result of these measurements is presented in Fig.~\ref{figeECCPMG} for the e$^-$ and in Fig.~\ref{fignECCPMG} for the $^{31}$P neutral \begin{suppfig}[!b]
\begin{center}
\includegraphics[width=1\textwidth]{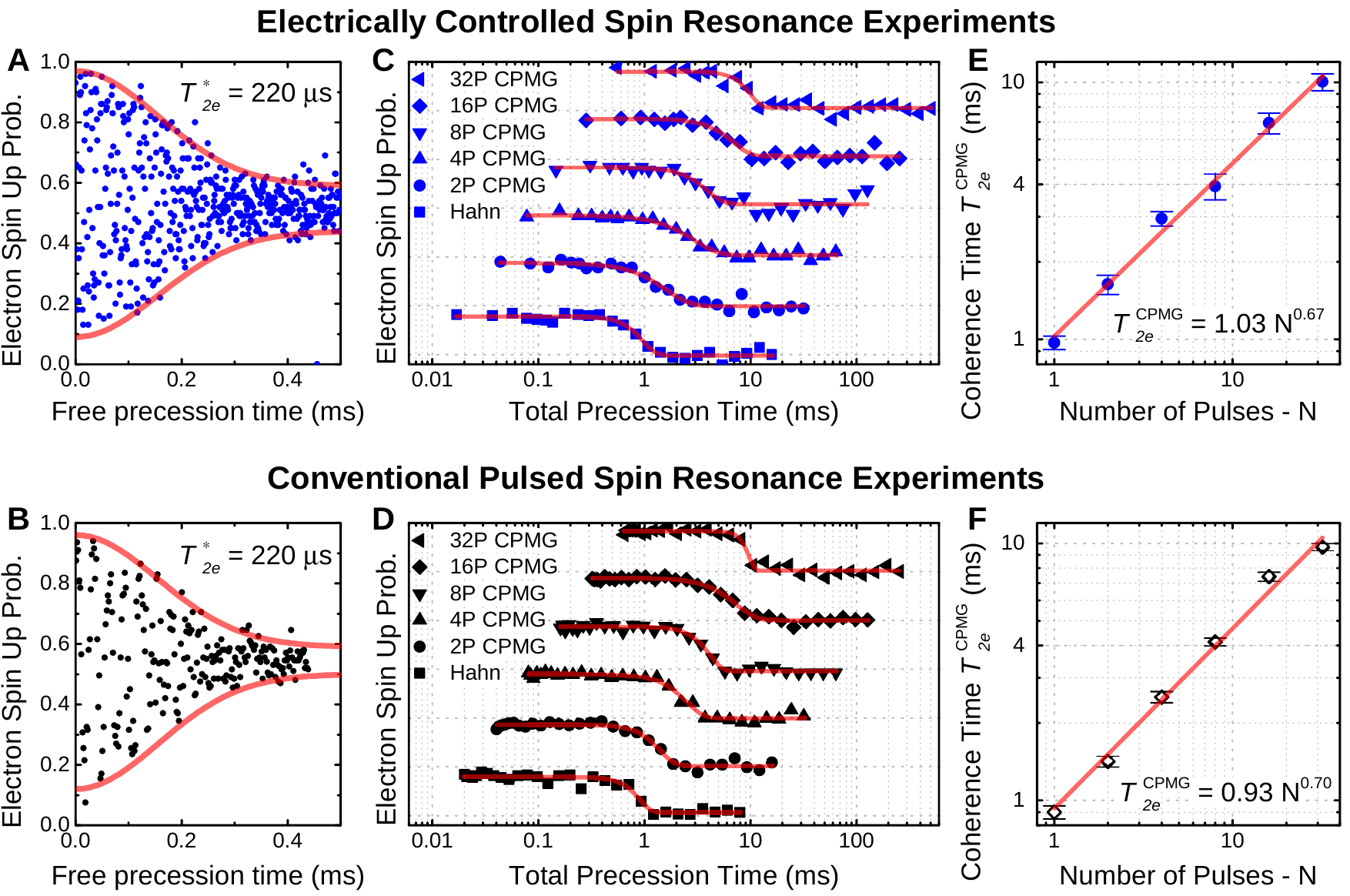}
\caption{\label{figeECCPMG} \textbf{Electrically controlled electron coherence times.}
\textbf{A,B} Ramsey experiments to extract $T_2^\ast$ for the electron qubit.
\textbf{C,D} CPMG dynamical decoupling decay traces for different numbers of refocussing pulses.
\textbf{E,F} Extracted coherence times $T^{\rm CMPG}_{2e}$ as a function of number of refocussing pulses $N$.}
\end{center}
\end{suppfig}
nucleus. Furthermore, the measurements on the e$^-$ have been performed as both electrically controlled and conventional, pulsed spin resonance experiments to directly compare these two measurements methods. The Ramsey experiments are displayed in Fig.~\ref{figeECCPMG}A,B, the measured CPMG decay traces and their fits are displayed in Fig.~\ref{figeECCPMG}C,D and Fig.~\ref{fignECCPMG}A, and the extracted decay times are plotted as a function of number of CPMG pulses in Fig.~\ref{figeECCPMG}E,F and Fig.~\ref{fignECCPMG}B.

The electron spin coherence times obtained for electrically controlled and pulsed experiments are identical within the error bars, indicating that there is no additional source of decoherence introduced by performing electrically-controlled measurements. The free induction decay time is $T_2^\ast = 220\mu s$ (see Fig.~\ref{figeECCPMG}A,B) and the extended coherence times reach values of $T_{2e}^{\rm CPMG}=10$~ms for 32 decoupling pulses (see Fig.~\ref{figeECCPMG}C-F) for both experimental methods.
The electron spin coherence times can be fitted with $T_{2e}^{\rm CPMG} \propto N^{\alpha/(\alpha+1)} = N^{0.67}$, where $\alpha = 2.0$ gives information about a colored noise spectrum with $S(\omega) \propto 1/\omega^{2.0}$ in good agreement with the data obtained by noise spectroscopy with pulsed-MW control sequences with  $T^{\rm pulse CPMG}_{2e} = 0.93 N^{0.70}$ ($\alpha = 2.3$). This is another strong indicator, that the electrically-controlled measurements do not introduce significant additional noise.
\begin{suppfig}[!t]
\begin{center}
\includegraphics[width=1\textwidth]{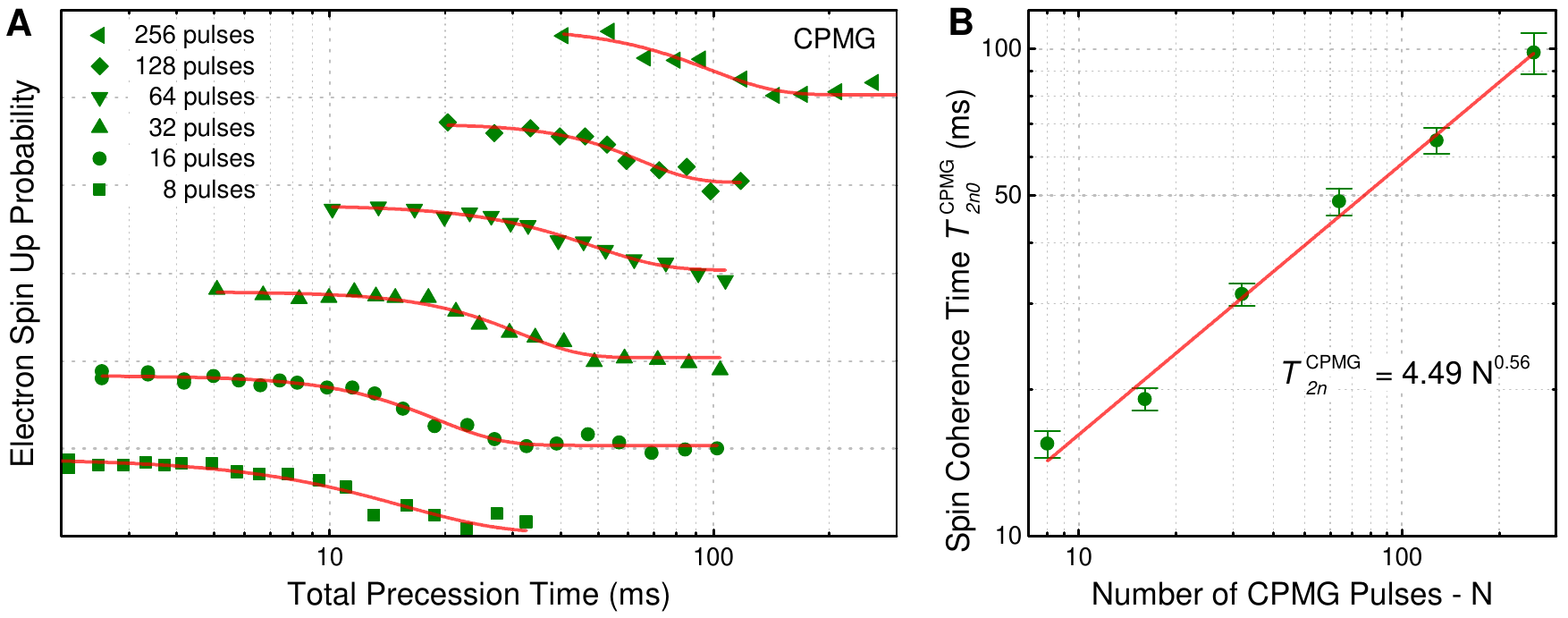}
\caption{\label{fignECCPMG} \textbf{Electrically controlled nuclear coherence times.}
\textbf{A} Set of EC CPMG dynamical decoupling performed on the $^{31}$P neutral nucleus.
\textbf{B} Extracted coherence times $T^{\rm CPMG}_{2n}$ as a function of number of decoupling pulses.}
\end{center}
\end{suppfig}

The coherence times of the neutral nucleus can be fitted with $T_{2n}^{\rm CPMG} \propto N^{\alpha/(\alpha+1)} = N^{0.56}$, with $\alpha = 1.3$ that hints towards $1/f$-type noise. In contrast to the measurements presented here, in measurements with pulse-RF control sequences we were not able to significantly extend $T_{2n}^{\rm CPMG}$ with dynamical decoupling sequences (see Ref.~\cite{Muhonen2014}). We believe that this is caused by a sensitivity of the donor system to the RF radiation, possibly by a process as simple as heating. For the pulse-RF technique, the RF power was chosen higher and pulsing of the RF source will lead to a non-equilibrium state. On the other hand for the electrically-controlled technique, the RF power was chosen lower (to reduce power broadening of the NMR transition) and the CW RF drive was leading to a steady-state situation.

\newpage
\bibliography{Papers}